\documentclass[10pt,twocolumn,letterpaper]{article}

\usepackage{cvpr}                %

\usepackage{newtxtext}
\usepackage{makecell}
\usepackage{multirow}
\usepackage{placeins}
\usepackage{float}
\usepackage{needspace}    %
\usepackage{adjustbox}     %

\newcommand{\best}[1]{\textbf{\boldmath #1}}
\newcommand{\nd}[1]{\underline{#1}}

\newcommand{\MambaSplit}{\textsc{MambaSplit}\xspace}
\newcommand{\UNet}{U-Net\xspace}
\newcommand{\muSplit}{$\mu$Split\xspace}
\newcommand{\denoiSplit}{denoiSplit\xspace}
\newcommand{\MicroSplit}{MicroSplit\xspace}
\newcommand{\scSplit}{scSplit\xspace}

\makeatletter
\newcommand{\suplabel}[2]{%
  \def\@currentlabel{#2}%
  \def\cref@currentlabel{[section][#2][]#2}%
  \label{#1}}
\makeatother
\newcommand{\supref}[1]{Supp.~Section~\ref{#1}}
\newcommand{\suppcref}[1]{Supp.~\cref{#1}}
\AtBeginDocument{%
  \crefname{figure}{Figure}{Figures}\Crefname{figure}{Figure}{Figures}%
  \crefname{table}{Table}{Tables}\Crefname{table}{Table}{Tables}%
  \crefname{section}{Section}{Sections}\Crefname{section}{Section}{Sections}%
}

\renewcommand{\topfraction}{0.9}
\renewcommand{\dbltopfraction}{0.9}
\renewcommand{\bottomfraction}{0.8}
\renewcommand{\textfraction}{0.07}
\renewcommand{\floatpagefraction}{0.75}
\renewcommand{\dblfloatpagefraction}{0.75}
\newif\ifshowchanges \showchangesfalse
\DeclareRobustCommand{\yhl}[1]{\bgroup\markoverwith{\textcolor{yellow!65}{\rule[-.55ex]{2pt}{2.45ex}}}\ULon{#1}}

\usepackage[normalem]{ulem}%
\definecolor{coedteal}{RGB}{0,140,140}

\makeatletter
\AddToShipoutPicture{\ifnum\c@page>1 \AtTextUpperLeft{\put(0,\LenToUnit{0.55cm}){\makebox[\textwidth][s]{\small
  \ifodd\c@page \hfill\MambaSplit\hspace{2em}\thepage \else \thepage\hspace{2em}A.~Kukde et al.\hfill \fi}}}\fi}
\makeatother

\newcommand{\figTeaser}{%
\begin{figure*}[t]\centering
\includegraphics[width=0.93\textwidth]{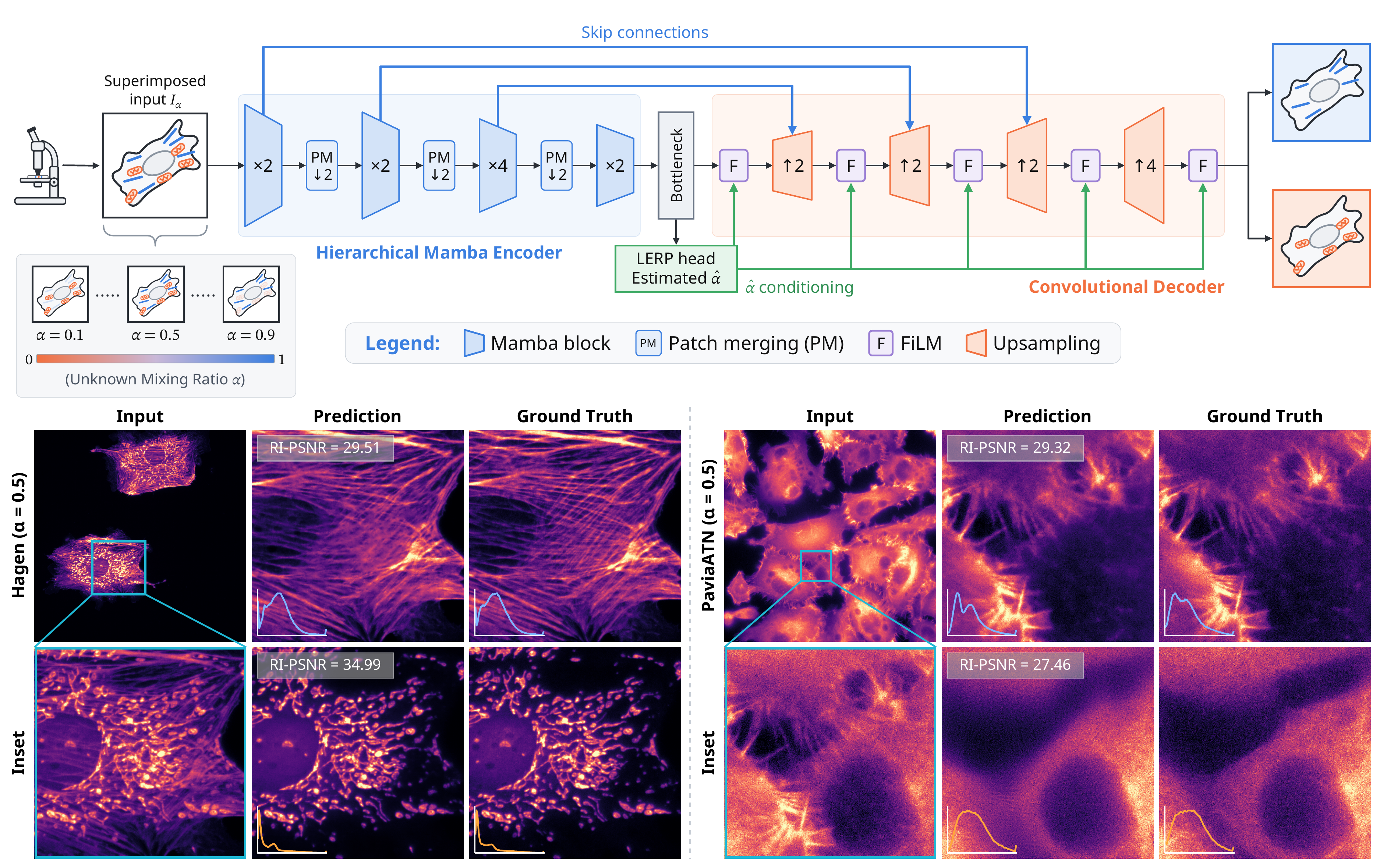}
\caption{\textbf{Semantic unmixing under unknown mixing severity.}
\emph{Top:}~\MambaSplit.
A superimposed input $I_\alpha$, the sum of two structures weighted by an unknown $\alpha$ and $1-\alpha$, is encoded by a hierarchical Mamba encoder with patch merging.
A LERP head estimates $\hat{\alpha}$ from the bottleneck, and FiLM layers condition every stage of the convolutional decoder on it, which predicts both structures in a single pass.
\emph{Bottom:}~Representative \MambaSplit predictions on one superimposed image each from the Hagen and PaviaATN datasets ($\alpha=0.5$, equally weighted standardized channels), shown as a crop (top left) and a magnified inset (bottom left), with RI-PSNR (dB) and intensity histograms of prediction and ground truth.}
\label{fig:teaser}
\end{figure*}
}

\newcommand{\figPatchSize}{%
\begin{figure}[t]\centering
\includegraphics[width=.8\columnwidth]{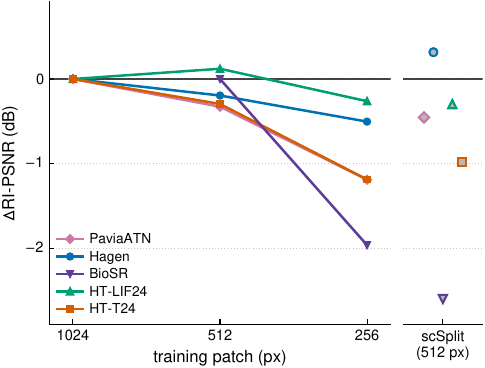}
\caption{\textbf{Effect of the \MambaSplit training patch size at a fixed pixel budget.} Change of RI-PSNR relative to the
$1024$-pixel model ($512$ on BioSR), averaged over nine mixing ratios; all configurations see the same pixels per training step (batch sizes 5, 20 and 80 at 1024, 512 and 256 pixels; 5 and 20 at 512 and 256 on BioSR; \suppcref{tab:patch-size-pixmatch}). Right: scSplit at its $512$-pixel tile against the same reference.}
\label{fig:patch-size}
\end{figure}
}

\newcommand{\figWeakStructureQual}{%
\begin{figure*}[!t]
\centering
\includegraphics[width=.9\textwidth]{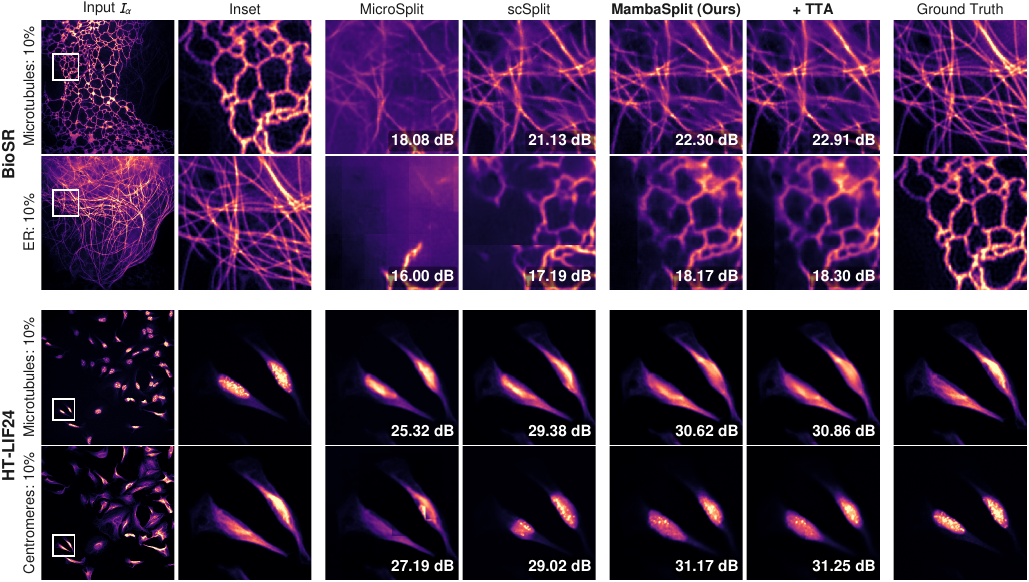}
\caption{\textbf{Qualitative results obtained when one structure is weak (dim) compared to the other.} Reconstruction of structures contributing only $10\%$ of the summed input on BioSR (microtubules, ER) and HT-LIF24 (microtubules, centromeres); row labels name the weak structure. Values report RI-PSNR (dB) on the displayed inset; $+$\,TTA is \MambaSplit{} with eight-fold test-time augmentation. For experimentally acquired superpositions, see \cref{fig:acquired-qual}.}
\label{fig:weak-structure-qual}
\end{figure*}
}

\newcommand{\figAcqQual}{%
\begin{figure*}[!t]
\centering
\includegraphics[width=.9\textwidth]{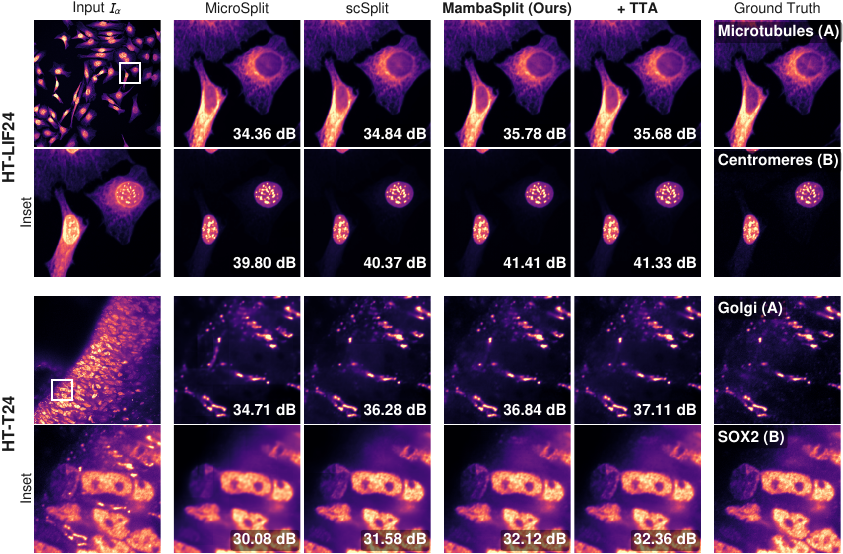}
\caption{\textbf{Qualitative results on acquired superpositions, directly imaged at the microscope.} Representative predictions from the best performing model of each method on HT-LIF24 (A: microtubules, B: centromeres) and HT-T24 (A: Golgi, B: SOX2-labeled nuclei); the upper row of each dataset shows structure A, the inset row structure B in the boxed region. Values report crop RI-PSNR (dB); $+$\,TTA is \MambaSplit{} with eight-fold test-time augmentation.}
\label{fig:acquired-qual}
\end{figure*}
}

\definecolor{cvprblue}{rgb}{0.21,0.49,0.74}
\usepackage[pagebackref,breaklinks,colorlinks,allcolors=cvprblue]{hyperref}
\usepackage{orcidlink}  %

\def\confName{CVPR}
\def\confYear{2027}

\title{MambaSplit: Severity-Robust and Inference-Efficient\\Semantic Unmixing of Fluorescence Microscopy Image Data}

\author{
Aman Kukde$^{1}$\,\orcidlink{0009-0008-3955-8382} \quad Ruggiero Santeramo$^{1}$\,\orcidlink{0000-0001-7107-235X} \quad Federico Carrara$^{1}$\,\orcidlink{0009-0005-9445-8497} \\
Ashesh Ashesh$^{1,2}$\,\orcidlink{0000-0003-3778-0576} \quad Igor Zubarev$^{1}$\,\orcidlink{0009-0008-6705-0949} \quad Florian Jug$^{1}$\,\orcidlink{0000-0002-8499-5812} \\[4pt]
$^{1}$Fondazione Human Technopole, Milan, Italy \qquad
$^{2}$IIT Hyderabad, Hyderabad, India \\
{\tt\small \{aman.kukde, florian.jug\}@fht.org}
}
\begin{document}
\maketitle

\begin{abstract}
The number of structures fluorescence microscopy can image in one sample is limited by how few fluorophores can be separated without significant spectral overlap.
Semantic unmixing circumvents this limit by imaging two structures in a single fluorescent channel and computationally separating the superimposed image into one image per structure, despite their complete overlap.
Because individual pixels rarely identify the structure they belong to, unmixing requires both local and long-range spatial context.
Additionally, the relative brightness of the two structures is unknown a priori and can be very uneven, so predictions must remain accurate across mixing severities, from balanced mixtures to those in which one structure is barely visible.
Existing methods either process small image patches or add context through auxiliary pathways that slow inference on large images.
We present \MambaSplit, a hierarchical state-space encoder with a convolutional decoder that captures context across large tiles in linear time and predicts both channels in a single deterministic pass.
\MambaSplit trains either on computationally summed channel images with random mixing ratios, where a lightweight head estimates the unknown ratio and conditions the decoder on it, or on experimentally acquired superpositions coming directly from a microscope.
Across five datasets and nine mixing ratios, \MambaSplit achieves the highest RI-PSNR, by up to 3.0\,dB, in 11 of 15 dataset--severity combinations and ties in one more, while running 5.9$\times$ faster than the strongest baseline.
On acquired superpositions, fine-tuning a model pretrained on summed inputs matches or exceeds all baselines.
Ablations show that accuracy and robustness come mainly from the training recipe: large tiles, mixing ratios sampled across their full range (up to 11\,dB over fixed-ratio training for weak and dominant structures), and independently drawn channel crops (3.9\,dB on BioSR).
The state-space backbone makes large tiles affordable, which will in turn enable high-content screening and event-driven smart microscopy applications.
\end{abstract}

\section{Introduction}
\label{sec:intro}

\figTeaser

Fluorescence microscopy is a key tool for biomedical research, but the number of biological structures it can image in one sample is limited.
Each structure needs its own fluorophore and detection channel, and because fluorophore emission spectra overlap, most experiments are restricted to a handful of channels~\cite{microsplit}.
Semantic unmixing~\cite{usplit,microsplit} lifts this limit computationally.
Several structures are imaged together in a single fluorescent channel, and a network unmixes the resulting superimposed image into one predicted channel image per structure.
The channels this frees up can be used for additional structures, or to save acquisition time and photon budget.

Unmixing is difficult because single pixels do not give away which structures contributed to the measured intensity.
Instead, this must be decided by the surrounding image content.
How much context is needed to make the best prediction is unknown a priori, but larger context has been shown to help~\cite{usplit}.
At the same time, the predicted channels must preserve thin filaments, sharp boundaries and other fine detail.
Unmixing therefore requires both local and long-range spatial context, on microscopy images that are often many megapixels in size.

A second difficulty is that the structures rarely contribute equally.
Their relative brightness depends on labeling, fluorophores and imaging settings, so the mixing ratio is unknown at test time, and the weaker structure can be almost buried under the dominant one (\cref{fig:teaser}, bottom left).
A practical unmixing model must remain accurate across this range of mixing severities without being given the ratio.

Existing methods meet these requirements only in part.
Patch-based networks such as \UNet~\cite{unet} or \scSplit~\cite{scsplit} see only the context inside a small input patch.
\muSplit~\cite{usplit}, \denoiSplit~\cite{denoisplit} and \MicroSplit~\cite{microsplit} showed that context from outside the patch improves unmixing, but they add it through auxiliary pathways such as lateral contextualization, and the variational models average many posterior samples per tile.
These additions make inference on large images slow.
\scSplit~\cite{scsplit} was the first to address mixing severity directly, training across mixing ratios and conditioning two reconstruction networks on the output of a ratio regressor.

Here, we present \MambaSplit, which combines a training recipe for large-context, severity-robust unmixing with an efficient architecture: a hierarchical state-space encoder and a convolutional decoder.
Selective state-space blocks~\cite{mamba,visionmamba} propagate information along the token sequence in linear time, and patch merging~\cite{swin} lets deeper stages operate on coarser versions of the input.
Together they capture context across large tiles, while skip connections and the convolutional decoder restore fine detail.
\MambaSplit predicts both output channels in a single deterministic pass.
It can be trained in two ways:
$(i)$~on computationally summed channel images, with a random mixing ratio for every training patch and the two channel crops drawn independently of each other, where a lightweight LERP head on the encoder bottleneck estimates this ratio to condition the decoder via FiLM~\cite{film};
$(ii)$~on experimentally acquired superpositions, which carry no ratio labels, either from scratch without the LERP head and conditioning, or by fine-tuning a model pretrained on summed inputs.

We evaluate \MambaSplit on five fluorescence microscopy datasets and nine mixing ratios.
\MambaSplit achieves the highest range-invariant PSNR (RI-PSNR)~\cite{care,usplit} in $11$ of $15$ dataset--severity combinations and ties in one more (HT-LIF24 Balanced), with gains of up to $3.0$\,dB, and stays within $0.9$\,dB of the best baseline in the remaining three.
Our ablations show that most of this accuracy comes from the training recipe: large training patches, training across mixing ratios, and independently drawn channel crops, which on BioSR improve \MambaSplit by $3.9$\,dB and \scSplit by $2.4$\,dB.
The state-space backbone makes the required large tiles affordable, which speeds up inference and reduces tiling artifacts (\mbox{\suppcref{tab:seam-patch}}).

With regard to inference speed, \MambaSplit is $5.9\times$ faster than \scSplit and $164\times$ faster than \MicroSplit.
This enables time-critical applications in acquisition-heavy high-content screening~\cite{boutros2015,bray2016} and event-driven smart microscopy~\cite{shroff2024,mahecic2022,alvelid2022}, with the reduced tiling artifacts additionally supporting such applications.
To compare tiling artifacts between methods that use different tile sizes, we also introduce a common-grid variant of the fraction of rejected tests (FRT)~\cite{switi} (\cref{sec:exp:tiling}).

\section{Related Work}
\label{sec:related}

\paragraph{Image decomposition and spectral unmixing.}
Decomposing an image into constituent layers is a long-standing problem in computer vision.
The layers can combine additively, as in reflection removal~\cite{zhang2018reflection}, or multiplicatively, as reflectance and shading in intrinsic image decomposition~\cite{barrow1978,grosse2009}, and generic layer separation has been addressed with coupled deep image priors~\cite{doubledip}.
These layers typically differ in physical origin or image statistics, which the separation can exploit.
In semantic unmixing, the two layers add up, and both are fluorescence images formed by the same optics, detector and noise process, and they can only be told apart by the morphology of the structures they show.
Spectral unmixing~\cite{keshava2002,bioucasdias2012,zimmermann2005} addresses a related problem with additional measurements: each pixel is recorded in several spectral bands and modeled as a linear combination of endmember spectra, whose per-pixel abundances are estimated, for example, by least squares~\cite{heinz2001, Slawski2011-nt}, non-negative matrix factorization~\cite{Lee1999-zk, Pengo2013} or deep learning-based methods~\cite{palsson2018, Hong2022-fr, Ghosh2022-fm, lambdaSplit}.
In fluorescence microscopy, spectral unmixing separates fluorophores with overlapping emission spectra only when multiple detection channels are available.
Semantic unmixing removes this requirement by operating on a single channel; our LERP head estimates one mixing ratio per tile, derived from image content rather than spectra.

\paragraph{Semantic image unmixing.}
Fluorescence unmixing methods recover structure-specific channels from a single superimposed image using spatial and morphological priors. Convolutional architectures such as U-Net~\cite{unet} combine multi-scale features, but their patch-based processing limits the available spatial context. $\mu$Split~\cite{usplit} addressed this with lateral contextualization and a hierarchical variational architecture, incorporating information from a larger surrounding region at limited memory cost. denoiSplit~\cite{denoisplit} combined supervised splitting with denoising, while MicroSplit~\cite{microsplit} brought together contextual modeling, denoising, and uncertainty estimation. Performance can nevertheless degrade when structures contribute unequally to the mixture. scSplit~\cite{scsplit} addresses this with severity-aware reconstruction based on InDI~\cite{indi}: a dedicated regressor estimates the mixing ratio, which conditions two reconstruction networks together with mixing-ratio-aware normalization. \MambaSplit adopts from \scSplit the training across mixing ratios and the estimation of the ratio to condition the reconstruction, but integrates long-range context and ratio estimation within a single encoder-decoder, using FiLM~\cite{film} to condition the joint reconstruction of both structures.
Conditioning a restoration network on a degradation level, given or estimated, is established in image restoration, for example for noise levels~\cite{ffdnet}, continuous restoration strengths~\cite{adafm,cresmd} and super-resolution~\cite{srmd,ikc}.
\paragraph{Global context for image reconstruction.}
Semantic unmixing requires both fine local detail and broad spatial context: local morphology is needed to preserve thin structures and boundaries, while spatially distant evidence can help resolve locally ambiguous patterns. Convolutional networks capture local structure efficiently but have limited receptive fields, whereas self-attention provides global interactions at quadratic cost~\cite{vit}. Selective state-space models such as Mamba~\cite{mamba} offer long-range dependency modeling with linear sequence complexity, making them attractive for large microscopy images. Recent image-restoration models have adapted this idea to preserve local detail while expanding spatial context~\cite{mambair}.
Transformers with efficient attention, such as SwinIR~\cite{swinir} and Restormer~\cite{restormer}, likewise enlarge the receptive field for restoration.
We build on the hierarchical design of Orochi~\cite{orochi}, which combines Mamba blocks with Swin-style patch merging~\cite{swin} for efficient multi-scale processing of biomedical images.
U-shaped Mamba models for biomedical segmentation~\cite{umamba,segmamba,swinumamba} and multi-directional scans~\cite{vmamba} pursue related designs.
We adopt this encoder hierarchy with a single forward scan and pair it with a LERP head that estimates the mixing ratio from the bottleneck and a FiLM-conditioned convolutional decoder that predicts both structures (\cref{sec:method:backbone}).

\section{MambaSplit}
\label{sec:method}

\subsection{Problem formulation and input construction}
\label{sec:method:problem}

Semantic image unmixing in fluorescence microscopy aims to recover the individual contributions of multiple biological structures from a single superimposed image~\cite{usplit}.
In this work, we focus on the two-structure setting, where the goal is to reconstruct two structure-specific channels, $A\in\mathbb{R}^{H\times W}$ and $B\in\mathbb{R}^{H\times W}$, from their mixture $\alpha A+(1-\alpha)B$.
For computationally summed inputs, we independently standardize each channel using its training-set statistics and construct the network's input as
\begin{equation}
I_\alpha=\alpha A_z+(1-\alpha)B_z,
\qquad
\alpha\in[0,1].
\label{eq:mix}
\end{equation}
Here, $A_z$ and $B_z$ denote the standardized channels, while $\alpha$ controls their relative contributions.
Mixtures near either extreme make the weaker component particularly difficult to recover.
Given \(I_\alpha\), \MambaSplit reconstructs the standardized channels \(A_z\) and \(B_z\), yielding predictions \(\hat{A}_z\) and \(\hat{B}_z\).
The mixing ratio is known when constructing training mixtures but unknown at inference.
Summed inputs are convenient because any two separately imaged structures yield training data with exact ground truth at every mixing ratio.
In an actual experiment, however, both structures are labeled in the same sample and imaged together in one channel, so the microscope records the superposition directly.
Such experimentally acquired superpositions are the real use case of semantic unmixing.
Image formation is still linear in the emitted fluorescence, but acquired superpositions carry no ratio labels, and their noise, background and alignment differ from those of a sum of separately acquired images.
\cref{sec:method:acquired} describes how \MambaSplit adapts to them.
\subsection{Architecture}
\label{sec:method:backbone}

\MambaSplit consists of a hierarchical state-space encoder, a mixing-ratio estimation head, and a shared convolutional decoder (\cref{fig:teaser}, top; full architecture in \suppcref{fig:arch-main}).
We call the estimation head the LERP head, because \cref{eq:mix} is a linear interpolation (LERP) between the two standardized channels with weight $\alpha$.
The encoder follows the hierarchical design of Orochi~\cite{orochi}: non-overlapping $4\times4$ patches are processed through four stages with $2\times2$ patch merging~\cite{swin}, and each block contains a selective state-space layer~\cite{mamba}.
This layer treats the patch tokens as one sequence, which it scans once in row-major order, that is, row by row starting from the top-left patch.

The LERP head predicts a mixing ratio \(\hat{\alpha}\) from the bottleneck features. FiLM layers~\cite{film} then use this estimate to condition the bottleneck and decoder features with channel-wise scale and bias parameters, allowing the reconstruction network to adapt to the predicted mixing severity. The conditioning design is detailed in \supref{sec:supp:archcond}, with ablations in \suppcref{tab:ratio-conditioning-ablation}.

The decoder then progressively upsamples the conditioned bottleneck representation, combines it with encoder features through skip connections, and jointly reconstructs both output channels.

\subsection{Training and inference}
\label{sec:method:training}

For each training patch, we sample $\alpha$ uniformly from $[0,1]$ and form the mixture as described in \cref{eq:mix}.
The crops of the two channels are drawn independently of each other rather than from the same index, which on BioSR improves RI-PSNR by $3.9$\,dB for \MambaSplit and by $2.4$\,dB for \scSplit (\supref{sec:supp:pairing}).
The training loss combines a Huber loss~\cite{huber1964} on each reconstructed channel with a squared-error term on the estimated ratio,
\begin{equation}
\mathcal{L}=
\mathrm{Huber}_{\delta}(\hat{A}_z,A_z)
+
\mathrm{Huber}_{\delta}(\hat{B}_z,B_z)
+
\lambda(\hat{\alpha}-\alpha)^2,
\label{eq:loss}
\end{equation}
with $\delta=1$ and $\lambda=0.1$.
The Huber loss is quadratic for residuals smaller than $\delta$ and linear beyond, so it behaves like a squared error on small residuals while limiting the influence of large ones.
During training, the decoder is conditioned on the ground-truth $\alpha$, while the ratio term trains the LERP head for inference.
At test time, the decoder instead receives the predicted $\hat{\alpha}$ and outputs both channels in one deterministic pass per tile.
More training and inference details are provided in \supref{sec:supp:training}.

\subsection{Adaptation to experimentally acquired superpositions}
\label{sec:method:acquired}

Acquired superpositions provide no ratio labels, so we consider two adaptation strategies.
Training from scratch uses an unconditioned \MambaSplit variant without the LERP head, FiLM conditioning and ratio term, which maps the acquired superposition directly to the two channels.
Fine-tuning starts from the model trained on summed inputs, freezes its LERP head, and uses the predicted ratio to condition the decoder through trainable FiLM layers while the reconstruction network is fine-tuned on acquired training pairs.
Removing the LERP head and FiLM layers before fine-tuning, instead of freezing the head, changes accuracy by at most $0.05$\,dB (\mbox{\suppcref{tab:finetune-variants}}), so the gain from fine-tuning comes from the summed-input initialization.
Acquired superpositions are standardized with their own training-set statistics, while targets keep their channel-specific standardization.
Results are reported in \cref{sec:exp:acquired}.

\section{Experiments and Results}
\label{sec:exp}

\subsection{Setup}
\label{sec:exp:setup}

\paragraph{Datasets.}
We evaluate on five benchmarks: Hagen~\cite{hagen}, BioSR~\cite{biosr}, PaviaATN~\cite{usplit}, and HT-T24 and HT-LIF24~\cite{microsplit}, using the unmixing tasks and data splits of scSplit~\cite{scsplit} (\suppcref{tab:datasets}). Image sizes range from $1004^2$ to $2720^2$. Unless stated otherwise, \MambaSplit{} is trained on $1024\times1024$ crops; for BioSR, whose native frames are $1004\times1004$, we use $512\times512$ crops. HT-T24 and HT-LIF24 additionally provide experimentally acquired superpositions with aligned component references, which we use to evaluate transfer and adaptation; their test sets contain $36$ and $10$ acquired superpositions, respectively.

\paragraph{Baselines.}
We compare against U-Net~\cite{unet}, $\mu$Split~\cite{usplit}, denoiSplit~\cite{denoisplit}, MicroSplit~\cite{microsplit}, InDI~\cite{indi}, and scSplit~\cite{scsplit}. Unless marked otherwise, results in \cref{tab:summed-severity-results} are taken from scSplit~\cite{scsplit}; for $\mu$Split, we use the strongest released architecture for each dataset. The reported Hagen and PaviaATN results use clipped targets, which discard intensity differences among the brightest structures.
We therefore evaluate on unclipped targets (\supref{sec:supp:targets}), retrain the affected baselines using their original recipes, and mark these results with a $^{\dagger}$; the published results are listed for reference.
Checkpoint selection is described in \supref{sec:supp:training}.

\paragraph{Metrics and regimes.}
The released benchmarks define nine mixing weights $w$ on raw intensities, whereas our $\alpha$ weights standardized channels; both inputs are identical up to an affine intensity transform. We evaluate \MambaSplit{} at the corresponding $\alpha$ for each $w$ (\supref{sec:supp:axis}), so all methods are evaluated on equivalent mixtures. Each output channel is assigned to a regime according to its contribution: Weak $\{0.1,0.2,0.3\}$, Balanced $\{0.4,0.5,0.6\}$, or Dominant $\{0.7,0.8,0.9\}$. Our main metric is RI-PSNR~\cite{care,usplit}, which removes global linear intensity differences before measuring reconstruction error. We also report MS-SSIM~\cite{msssim}. LPIPS~\cite{lpips} and PSNR are reported for all methods in \suppcref{tab:extended-metrics}.

\subsection{Reconstruction from summed inputs}
\label{sec:exp:summed-results}

\begin{table}[!t]
\centering
\caption{\textbf{Quantitative results across mixing severities.} Range-invariant PSNR (RI-PSNR, dB) and MS-SSIM in the Dominant (D), Balanced (B), and Weak (W) regimes. Each output channel is assigned to a regime by its own weight in the mixture: Dominant for weights $\{0.7,0.8,0.9\}$, Balanced for $\{0.4,0.5,0.6\}$, and Weak for $\{0.1,0.2,0.3\}$ (\cref{sec:exp:setup}). Rows marked with a $^{\dagger}$ are our retraining on raw unclipped data; on PaviaATN and Hagen, the published results above them were scored against clipped targets and are not ranked; TTA denotes test-time augmentation. Within each dataset and metric, best and second-best results are \best{bold} and \nd{underlined}, respectively.}
\label{tab:summed-severity-results}
\footnotesize\setlength{\tabcolsep}{3.5pt}\renewcommand{\arraystretch}{0.79}
\begin{adjustbox}{max width=\columnwidth}
\begin{tabular}{@{}c l ccc@{\hspace{1.5em}}ccc@{}}
\toprule
\multirow{2}{*}{\textbf{Data}} & \multirow{2}{*}{\textbf{Model}} & \multicolumn{3}{c}{\textbf{RI-PSNR (dB)}} & \multicolumn{3}{c}{\textbf{MS-SSIM}} \\
\cmidrule(lr){3-5}\cmidrule(l){6-8}
 & & D & B & W & D & B & W \\
\midrule
\multirow{11}{*}{\rotatebox[origin=c]{90}{\textbf{PaviaATN}}}
 & U-Net & 29.3 & 25.4 & 21.2 & 0.870 & 0.743 & 0.568 \\
 & $\mu$Split & 27.9 & 25.2 & 21.3 & 0.908 & 0.808 & 0.648 \\
 & denoiSplit & 27.3 & 26.2 & 21.8 & 0.857 & 0.843 & 0.750 \\
 & MicroSplit & 24.0 & 21.8 & 18.8 & 0.771 & 0.688 & 0.540 \\
 & InDI & 29.9 & 23.9 & 21.7 & 0.943 & 0.858 & 0.741 \\
 & scSplit & 35.1 & 27.6 & 24.3 & 0.977 & 0.907 & 0.823 \\
 \cmidrule(l){2-8}
 & U-Net$^{\dagger}$ & 30.7 & 26.9 & 21.8 & 0.902 & 0.816 & 0.647 \\
 & MicroSplit$^{\dagger}$ & 26.1 & 26.7 & 22.3 & 0.862 & 0.867 & 0.763 \\
 & scSplit$^{\dagger}$ & 39.4 & \nd{30.4} & 27.3 & 0.980 & \nd{0.918} & \nd{0.862} \\
\cmidrule(l){3-8}
 & \textbf{\MambaSplit{}} & \nd{40.1} & 30.2 & \nd{28.1} & \nd{0.981} & 0.915 & 0.859 \\
 & \quad $+$ TTA & \best{40.7} & \best{30.8} & \best{28.4} & \best{0.982} & \best{0.921} & \best{0.865} \\
\midrule
\multirow{11}{*}{\rotatebox[origin=c]{90}{\textbf{Hagen}}}
 & U-Net & 31.8 & 28.2 & 22.0 & 0.965 & 0.921 & 0.833 \\
 & $\mu$Split & 33.9 & 32.4 & 23.6 & 0.962 & 0.960 & 0.858 \\
 & denoiSplit & 32.4 & 31.9 & 23.1 & 0.958 & 0.954 & 0.851 \\
 & MicroSplit & 34.7 & 31.7 & 25.0 & 0.957 & 0.956 & 0.862 \\
 & InDI & 33.1 & 32.1 & 24.2 & 0.963 & 0.965 & 0.879 \\
 & scSplit & 40.9 & 33.9 & 29.3 & 0.994 & 0.977 & 0.934 \\
 \cmidrule(l){2-8}
 & U-Net$^{\dagger}$ & 43.8 & 39.5 & 31.6 & 0.971 & 0.976 & 0.914 \\
 & MicroSplit$^{\dagger}$ & 40.7 & 38.1 & 31.8 & \nd{0.974} & 0.980 & 0.921 \\
 & scSplit$^{\dagger}$ & \best{48.0} & \best{40.9} & 36.0 & \best{0.997} & \best{0.988} & \best{0.964} \\
\cmidrule(l){3-8}
 & \textbf{\MambaSplit{}} & 47.2 & \nd{40.5} & \nd{36.3} & \best{0.997} & \nd{0.987} & \nd{0.962} \\
 & \quad $+$ TTA & \nd{47.7} & \best{40.9} & \best{36.6} & \best{0.997} & \best{0.988} & \best{0.964} \\
\midrule
\multirow{8}{*}{\rotatebox[origin=c]{90}{\textbf{BioSR}}}
 & U-Net & 37.2 & 33.7 & 25.6 & 0.924 & 0.958 & 0.740 \\
 & $\mu$Split & 37.8 & 33.5 & 25.7 & 0.918 & 0.959 & 0.738 \\
 & denoiSplit & 36.4 & 33.1 & 25.3 & 0.929 & 0.957 & 0.733 \\
 & MicroSplit & 38.5 & 34.3 & 26.6 & 0.932 & 0.966 & 0.759 \\
 & InDI & 35.9 & 33.4 & 26.3 & 0.917 & 0.953 & 0.802 \\
 & scSplit & 40.1 & 35.3 & 28.7 & 0.987 & 0.973 & 0.889 \\
\cmidrule(l){3-8}
 & \textbf{\MambaSplit{}} & \nd{43.1} & \nd{37.7} & \nd{31.1} & \nd{0.993} & \nd{0.982} & \nd{0.922} \\
 & \quad $+$ TTA & \best{43.9} & \best{38.1} & \best{31.5} & \best{0.994} & \best{0.983} & \best{0.926} \\
\midrule
\multirow{8}{*}{\rotatebox[origin=c]{90}{\textbf{HT-LIF24}}}
 & U-Net & 45.9 & 44.6 & 36.0 & 0.980 & 0.986 & 0.939 \\
 & $\mu$Split & 46.7 & 45.1 & 36.6 & 0.978 & 0.986 & 0.940 \\
 & denoiSplit & 44.8 & 42.9 & 35.8 & 0.981 & 0.985 & 0.938 \\
 & MicroSplit & 45.0 & 43.7 & 36.5 & 0.982 & 0.986 & 0.939 \\
 & InDI & 45.2 & 43.9 & 37.6 & 0.976 & 0.991 & 0.963 \\
 & scSplit & 51.8 & \nd{45.5} & 39.9 & \nd{0.998} & \nd{0.994} & \nd{0.976} \\
\cmidrule(l){3-8}
 & \textbf{\MambaSplit{}} & \nd{52.3} & \nd{45.5} & \nd{40.3} & \best{0.999} & \best{0.996} & \best{0.988} \\
 & \quad $+$ TTA & \best{52.9} & \best{46.0} & \best{40.6} & \best{0.999} & \best{0.996} & \best{0.988} \\
\midrule
\multirow{8}{*}{\rotatebox[origin=c]{90}{\textbf{HT-T24}}}
 & U-Net & 37.9 & 37.5 & 30.1 & 0.963 & 0.965 & 0.883 \\
 & $\mu$Split & 37.6 & 36.9 & 29.9 & 0.954 & 0.959 & 0.880 \\
 & denoiSplit & 37.3 & 37.5 & 31.0 & 0.954 & 0.964 & 0.896 \\
 & MicroSplit & 36.9 & 36.6 & 30.3 & 0.950 & 0.959 & 0.891 \\
 & InDI & 37.6 & 36.5 & 30.5 & 0.962 & 0.966 & 0.909 \\
 & scSplit & 44.5 & 39.1 & 34.7 & \nd{0.995} & \nd{0.981} & \nd{0.943} \\
\cmidrule(l){3-8}
 & \textbf{\MambaSplit{}} & \nd{46.5} & \nd{39.8} & \nd{35.1} & \nd{0.995} & 0.980 & 0.942 \\
 & \quad $+$ TTA & \best{47.2} & \best{40.0} & \best{35.3} & \best{0.996} & \best{0.982} & \best{0.944} \\
\bottomrule
\end{tabular}
\end{adjustbox}
\end{table}

\cref{tab:summed-severity-results} compares methods across the three severity regimes, with the same \MambaSplit{} checkpoint evaluated at all nine ratios for each dataset. In RI-PSNR, single-pass \MambaSplit is best in $11$ of the $15$ dataset--severity combinations and ties scSplit on HT-LIF24 Balanced ($+0.09$\,dB; both $45.5$\,dB).
The exceptions are Hagen Dominant and Balanced, where scSplit$^{\dagger}$ leads by $0.83$ and $0.37$\,dB, and PaviaATN Balanced, where it leads by $0.21$\,dB.
Test-time augmentation (TTA; averaging predictions over the eight combinations of $90^\circ$ rotations and flips of the input, \supref{sec:supp:targets}) adds a further $0.2$--$0.7$\,dB.
The largest gains are on BioSR, where \MambaSplit{} outperforms every baseline by at least $2.4$\,dB in all three regimes ($2.4$--$3.0$\,dB over the strongest baseline, the released scSplit).
With the same independent sampling, scSplit comes within $0.2$\,dB on average (\supref{sec:supp:pairing}).

In MS-SSIM, single-pass \MambaSplit is best or tied in $9$ of the $15$ combinations and within $0.003$ of the best in the remaining six.
Per-ratio results are reported in \suppcref{tab:perratio-a,tab:perratio-b}.

\cref{fig:weak-structure-qual} shows reconstruction of the weakly represented structure at mixing weight $w=0.1$ on BioSR and HT-LIF24. \MambaSplit{} better preserves weak structures under such severe imbalance. On BioSR, MicroSplit loses much of the weak network structure, while on HT-LIF24 it leaves residual signal from the dominant component. Additional representative weak- and dominant-component examples for all datasets are shown in Supp.~\crefrange{fig:gallery-PaviaATN}{fig:gallery-HTT24}.

\figWeakStructureQual

\subsection{Performance on experimentally acquired superpositions}
\label{sec:exp:acquired}

Models trained on computationally summed inputs transfer imperfectly
to experimentally acquired superpositions
(\cref{tab:acquired-results}, \mbox{\cref{fig:acquired-qual}}). On HT-T24, \MambaSplit{} and scSplit
reach $35.55$ and $35.97$\,dB, respectively; on HT-LIF24,
they reach $40.57$ and $40.90$\,dB.

Training on acquired superpositions with separately acquired
component references (Training Mode III of
MicroSplit~\cite{microsplit}) improves RI-PSNR across methods by
an average of 1.1\,dB on HT-T24 and 3.2\,dB on HT-LIF24.
Because acquired pairs lack supervised mixing ratios,
\MambaSplit{} is trained from scratch without the LERP head or
FiLM conditioning (\cref{sec:method:acquired}). This setting
also allows comparison with baselines trained directly on
acquired data.

We additionally fine-tune each method from its summed-input
checkpoint using half the original learning rate. Compared with
training from scratch on acquired data, fine-tuning improves
\MambaSplit{} by $0.71$\,dB on HT-LIF24 and
$0.38$\,dB on HT-T24, versus at most $0.41$ and
$0.21$\,dB for the baselines. \MambaSplit{} reaches
$44.28$\,dB on HT-LIF24, effectively matching scSplit
at $44.27$\,dB, and $37.34$\,dB on HT-T24,
the highest result among the evaluated methods.

\begin{table}[!t]
\centering
\small
\setlength{\tabcolsep}{3pt}
\caption{\textbf{Quantitative results on acquired superpositions.} RI-PSNR (dB) on experimentally acquired superpositions for three training settings: trained on computationally summed inputs only (Summed training), trained from scratch on acquired training pairs (Acquired training), and initialized from the summed-input model and fine-tuned on acquired training pairs to convergence (Acquired fine-tuning). Best results per dataset are shown in \textbf{bold}.}
\label{tab:acquired-results}
\begin{adjustbox}{max width=\columnwidth}
\begin{tabular}{@{}ll ccc@{}}
\toprule
\textbf{Dataset} & \textbf{Model}
& \makecell{Summed\\training}
& \makecell{Acquired\\training}
& \makecell{Acquired\\fine-tuning} \\
\midrule
HT-LIF24
    & MicroSplit    & 40.08 & 43.72 & 43.92 \\
    & scSplit       & 40.90 & 43.86 & 44.27 \\
    & \MambaSplit{} & 40.57 & 43.57 & \textbf{44.28} \\
\midrule
HT-T24
    & MicroSplit    & 35.77 & 36.55 & 36.76 \\
    & scSplit       & 35.97 & 37.01 & 37.15 \\
    & \MambaSplit{} & 35.55 & 36.96 & \textbf{37.34} \\
\bottomrule
\end{tabular}
\end{adjustbox}
\end{table}

\figAcqQual

\subsection{Inference cost}
\label{sec:exp:inference-cost}

\MambaSplit{} processes large tiles and predicts both channels
in a single forward pass, substantially reducing inference cost
(\cref{tab:inference-cost}). On a $2720\times2720$ PaviaATN
image, it requires 64 tile evaluations, compared with 1,000
for \denoiSplit{}, 2,100 for \scSplit{}, and 352,800 for
\MicroSplit{} under their published inference settings.
It takes $3.18$\,s per image, $5.9\times$ faster than
\scSplit{} and $164\times$ faster than \MicroSplit{}.
The batching policy and profiling protocol are detailed in
\supref{sec:supp:profiling}.

The U-Net baseline is faster and more memory-efficient,
but substantially less accurate: \MambaSplit{} achieves
4.7\,dB higher RI-PSNR on average across the five datasets
and three severity regimes. Our training recipe also transfers to U-Nets, achieving comparable accuracy with a capacity-matched model, but requiring $3.3\times$ the inference time and $2\times$ the peak GPU memory \mbox{(\suppcref{tab:unet-conditioning})}.

\begin{table}[!t]\centering
\caption{\textbf{Full-frame inference cost and reconstruction quality.} For one $2720\!\times\!2720$ PaviaATN image, with each method's published inference settings (\supref{sec:supp:profiling}), timed on the same $71$\,GB GPU with each method at their largest allowable batch size. For peak memory and single tile see \mbox{\suppcref{tab:batch-policy}}. RI-PSNR is averaged over both output channels and the nine released mixing ratios $w\in\{0.1,0.2,\dots,0.9\}$. Rows marked with a $^{\dagger}$ were retrained on raw unclipped data, to their advantage.}
\label{tab:inference-cost}
\small\setlength{\tabcolsep}{3pt}\renewcommand{\arraystretch}{1.05}
\begin{adjustbox}{max width=\columnwidth}
\begin{tabular}{@{}l c c r r c@{}}
\toprule
\textbf{Method} & \textbf{Tile} & \makecell{\textbf{Par.}\\\textbf{(M)}}
& \textbf{Batch} & \makecell{\textbf{Time}\\\textbf{(s)}} & \makecell{\textbf{RI-PSNR}\\\textbf{(dB)}} \\
\midrule
U-Net$^{\dagger}$ & $512\!\times\!512$  & 17.30 & 100 & 0.76   & 26.46 \\
denoiSplit & $512\!\times\!512$  & 3.25  & 64 & 24.85  & 25.14 \\
MicroSplit$^{\dagger}$ & $64\!\times\!64$   & 3.57  & 1024 & 522.36 & 25.05 \\
scSplit$^{\dagger}$ & $512\!\times\!512$  & 7.30  & 128 & 18.69  & 32.36 \\
\midrule
\MambaSplit{} & $1024\!\times\!1024$ & 36.41 & 64 & 3.18  & 32.81 \\
\bottomrule
\end{tabular}
\end{adjustbox}
\end{table}

\subsection{Tiling artifacts}
\label{sec:exp:tiling}

We quantify tiling artifacts with the \emph{Fraction of Rejected Tests} (FRT) of~\cite{switi}: per tile, a permutation test compares intensity gradients across the seams with nearby control gradients, and FRT is the fraction of tests in which the seams are statistically distinguishable from their neighborhood.
FRT evaluates each method on its own tiling grid; the common-grid FRT (cFRT), which we introduce here, evaluates all methods on the same 256\,px seam grid and on the interior tiles they share (protocol in \supref{sec:supp:tiling}).

\MambaSplit with TTA achieves the best FRT on HT-T24, the second-best FRT on HT-LIF24, and the second-best cFRT on PaviaATN, HT-LIF24 and HT-T24 (\suppcref{tab:seam-balanced}).
Otherwise, the best scores mostly go to \UNet and \muSplit, which keep only the centre of each predicted tile (inner tiling~\cite{usplit,switi}) but reach considerably lower reconstruction quality (RI-PSNR).
On Hagen and BioSR, \MambaSplit has higher seam scores than \UNet, possibly because neighboring tiles receive different ratio estimates.
Over-smoothed outputs can suppress seams without improving unmixing (\suppcref{fig:seam-quality}), so FRT and cFRT must be read together with reconstruction metrics (here RI-PSNR).

\subsection{Ablations}
\label{sec:exp:ablation}

\paragraph{Diverse-ratio training and conditioning.}
\cref{tab:recipe-ablation} shows that robustness to mixing severity comes primarily from training across ratios.
Relative to training at a fixed $\alpha=0.5$, uniform ratio sampling improves the Dominant (D) and Weak (W) regimes by $3.60$--$11.10$\,dB per dataset when neither variant uses FiLM (\suppcref{tab:ratio-conditioning-ablation}; averages in \cref{tab:recipe-ablation}).
FiLM conditioning provides a smaller additional gain, while access to the true ratio at inference improves performance by up to $0.82$\,dB (BioSR).
Per-dataset results and ratio-estimation accuracy are reported in \suppcref{tab:ratio-conditioning-ablation} and \supref{sec:supp:ratio}.

\begin{table}[!t]
\centering
\small
\setlength{\tabcolsep}{3pt}
\caption{\textbf{Ablation studies.} RI-PSNR (dB) averaged over all five datasets. \MambaSplit reports absolute performance; the remaining rows report changes relative to \MambaSplit in the Dominant (D), Balanced (B) and Weak (W) regimes (\cref{sec:exp:setup}).}
\label{tab:recipe-ablation}
\begin{adjustbox}{max width=\columnwidth}
\begin{tabular}{@{}l ccc@{}}
\toprule
\textbf{Variant} & D & B & W \\
\midrule
\MambaSplit{} & 45.85 & 38.76 & 34.15 \\
\midrule
$-$FiLM & $-$0.36 & $-$0.24 & $-$0.12 \\
trained with $\alpha=0.5$ & $-$8.24 & $-$2.77 & $-$5.11 \\
trained with $\alpha=0.5$: $-$FiLM & $-$8.29 & $-$2.74 & $-$5.14 \\
\midrule
\MambaSplit{}: true $\alpha$ at inference & +0.29 & +0.19 & +0.19 \\
\MambaSplit{}: TTA at inference & +0.63 & +0.41 & +0.33 \\
\bottomrule
\end{tabular}
\end{adjustbox}
\end{table}

\paragraph{Training patch size.}
At a fixed pixel budget per training step, 512- and 1024-pixel patches differ by at most $0.3$\,dB, whereas 256-pixel patches lose up to $2.0$\,dB (\cref{fig:patch-size}; \suppcref{tab:patch-size-pixmatch}), consistent with a benefit of spatial context. We use 1024-pixel tiles ($512$ on BioSR) because, at the same relative overlap, they process a frame ${\sim}1.5\times$ faster than 512-pixel tiles (\suppcref{tab:tile-speed}).

\figPatchSize

\section{Discussion and limitations}
\label{sec:discussion}

\MambaSplit reaches the highest RI-PSNR in $11$ of $15$ dataset--severity combinations and ties in one more, requiring only $64$ deterministic evaluations of larger tiles, against $352{,}800$ for the posterior averaging of \MicroSplit (\cref{sec:exp:inference-cost}).
Our ablations attribute most of this accuracy to our training recipe (\cref{sec:exp:ablation}, \supref{sec:supp:pairing}), which also transfers to convolutional U-Nets (\suppcref{tab:unet-conditioning}).
\MambaSplit{}'s state-space backbone makes such large tiles affordable, $1.3$--$3.3\times$ faster than these U-Nets on the same tiles (\cref{sec:exp:inference-cost}, \mbox{\suppcref{tab:unet-conditioning}}), which matters for time-critical applications such as high-content screening and event-driven smart microscopy.

\paragraph{Severity robustness comes through training.}
The ablations in \cref{sec:exp:ablation} show that training across mixing ratios is what makes a single network robust to mixing severity, and that ratio conditioning adds a further $0.1$--$0.4$\,dB on average (\cref{tab:recipe-ablation}).
The LERP head also returns a per-tile estimate of the mixing ratio, and the small gap between estimated and true ratio bounds what better ratio estimation could gain.

\paragraph{Acquired superpositions.}
Without adaptation, and on both datasets also when trained from scratch, \MambaSplit trails \scSplit on acquired data.
Only fine-tuning a model pretrained on summed inputs makes it match or exceed all baselines on both datasets on average, and exceed them in three of four individual channels (\cref{sec:exp:acquired}, \suppcref{tab:acquired-per-channel}), so summed-input pretraining matters beyond its role as a benchmark, also for real-world applications.

\paragraph{Limitations.}
\label{sec:limitations}
\MambaSplit, like the ratio parameterization of \cref{eq:mix}, assumes two structures per channel; more structures require a vector of mixing weights.
As a deterministic model, it provides no uncertainty estimates, unlike the posterior samples of \muSplit, \denoiSplit and \MicroSplit, so plausible but wrong structures cannot be flagged, which matters when predictions feed into downstream biological analysis.
Training inputs follow a linear mixing model of standardized channels, and one ratio is estimated per tile, although relative brightness can vary within a tile.
Each Mamba block uses a single forward row-major scan, and we have not studied how scan order affects how context propagates to the output.

\section{Conclusion}
\label{sec:conclusion}

We presented \MambaSplit, which unmixes two fluorescent structures from large tiles in a single deterministic pass, fast enough to enable time-critical applications such as high-content screening and event-driven smart microscopy.
Across five datasets and nine mixing ratios, it achieves the best RI-PSNR in most dataset--severity combinations while running 5.9$\times$ faster than the strongest baseline, and after fine-tuning it matches or exceeds all baselines on acquired superpositions.
Most of this accuracy comes from the training recipe, and the state-space backbone makes large tiles affordable.
Natural next steps are unmixing more than two structures and adding calibrated uncertainty by combining state-space backbones with posterior models.
Code and trained models will be made publicly available under permissive licenses.

\section*{LLM Usage Statement}
We used an LLM-based assistant (Claude, Anthropic) for coding and writing.
\emph{Code and experiments:} writing and debugging training, evaluation and plotting scripts, managing cluster jobs, and cross-checking reported numbers against experiment outputs.
\emph{Figures:} plotting code and code for schematic diagrams; the microscopy images and predictions shown are unedited data and outputs of the evaluated models.
\emph{Text:} drafting and editing of text, grammar, literature search, and verification of bibliography entries.
\emph{Ideation:} suggestions for analyses, diagnostics and presentation; the authors chose the research questions, methods and experiments and verified all results.
The authors designed all experiments, verified all content, and take full responsibility for it.

{
    \small
    \bibliographystyle{ieeenat_fullname}
    \bibliography{references}
}

\setcounter{topnumber}{3}\setcounter{bottomnumber}{2}\setcounter{totalnumber}{5}
\setcounter{dbltopnumber}{3}
\renewcommand{\topfraction}{0.92}\renewcommand{\bottomfraction}{0.70}
\renewcommand{\dbltopfraction}{0.92}\renewcommand{\textfraction}{0.06}
\renewcommand{\floatpagefraction}{0.85}\renewcommand{\dblfloatpagefraction}{0.70}

\setcounter{table}{0}\renewcommand{\thetable}{S\arabic{table}}
\setcounter{figure}{0}\renewcommand{\thefigure}{S\arabic{figure}}

\twocolumn[
\begin{center}
    {\Large \textbf{\MambaSplit: Severity-Robust and Inference-Efficient\\Semantic Unmixing of Fluorescence Microscopy Image Data}}
\end{center}
\vspace{0.5em}
\begin{center}
    {\Large \textbf{\textit{Supplementary Material}}}
\end{center}
\vspace{1.0em}
]

\newcommand{\suppendpage}{\par\csname if@firstcolumn\endcsname\null\newpage\fi\null\newpage}

\newcommand{\figAcqPatchSize}{%
\begin{figure}[!htbp]\centering
\includegraphics[width=\columnwidth]{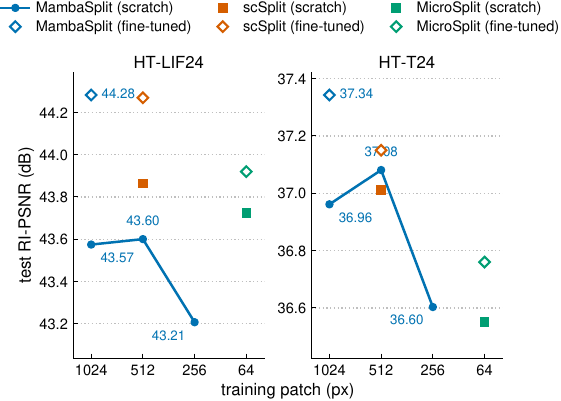}
\caption{\textbf{Patch size under acquired-input training.} Test RI-PSNR for \MambaSplit{} trained
from scratch on acquired pairs using $256$-, $512$- and $1024$-pixel patches, and the fine-tuned
$1024$-pixel model. scSplit and MicroSplit are placed at their own training patch ($512$ and $64$
pixels), from scratch (filled) and fine-tuned to convergence (open; \cref{tab:acquired-per-channel});
they are references, not part of the controlled patch-size ablation. Patch positions are
categorical, not to scale. All \MambaSplit{} patch sizes use inference stride $256$.
Each point is one training run, at the checkpoint with the lowest validation loss; the $1024$-pixel points are the checkpoints of
\cref{tab:acquired-results}. The $512$- and $256$-pixel runs train with the same number of pixels per step as the $1024$ model.}
\label{fig:acquired-patch-size}
\end{figure}
}

\newcommand{\figSynGalleries}{%
\begin{figure*}[tbp]\centering
\includegraphics[width=\textwidth]{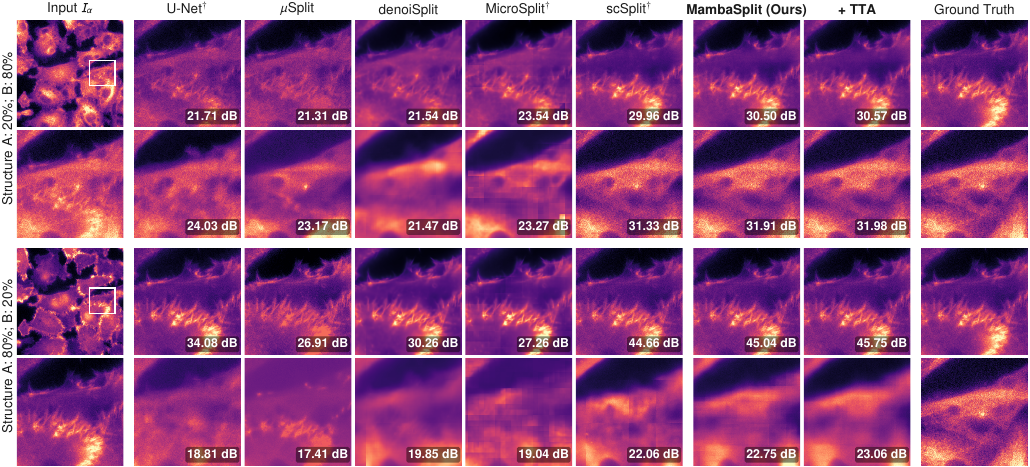}
\caption{\textbf{Weak- and dominant-component recovery on PaviaATN.} One test frame and crop at two mixing ratios: structure~A at $20\%$ and B at $80\%$ (top pair), and the reverse (bottom pair). Within each pair, the first row shows structure~A and the second structure~B. The input column shows the full frame with the crop marked, and the zoomed crop. Values report crop RI-PSNR (dB); $+$\,TTA is \MambaSplit{} with eight-view test-time augmentation. $^{\dagger}$ marks the unclipped retrains of \cref{tab:summed-severity-results}; the other baselines are the released models.}
\label{fig:gallery-PaviaATN}
\end{figure*}
\begin{figure*}[tbp]\centering
\includegraphics[width=\textwidth]{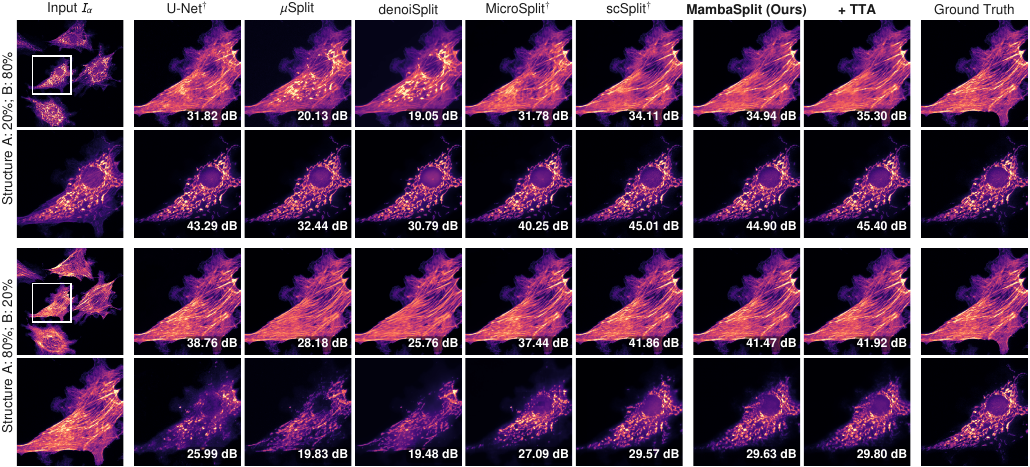}
\caption{\textbf{Weak- and dominant-component recovery on Hagen.} One test frame and crop at two mixing ratios: structure~A at $20\%$ and B at $80\%$ (top pair), and the reverse (bottom pair). Within each pair, the first row shows structure~A and the second structure~B. The input column shows the full frame with the crop marked, and the zoomed crop. Values report crop RI-PSNR (dB); $+$\,TTA is \MambaSplit{} with eight-view test-time augmentation. $^{\dagger}$ marks the unclipped retrains of \cref{tab:summed-severity-results}; the other baselines are the released models, and $\mu$Split and denoiSplit are scored against the clipped target, as in \cref{tab:summed-severity-results}.}
\label{fig:gallery-Hagen}
\end{figure*}
\begin{figure*}[tbp]\centering
\includegraphics[width=\textwidth]{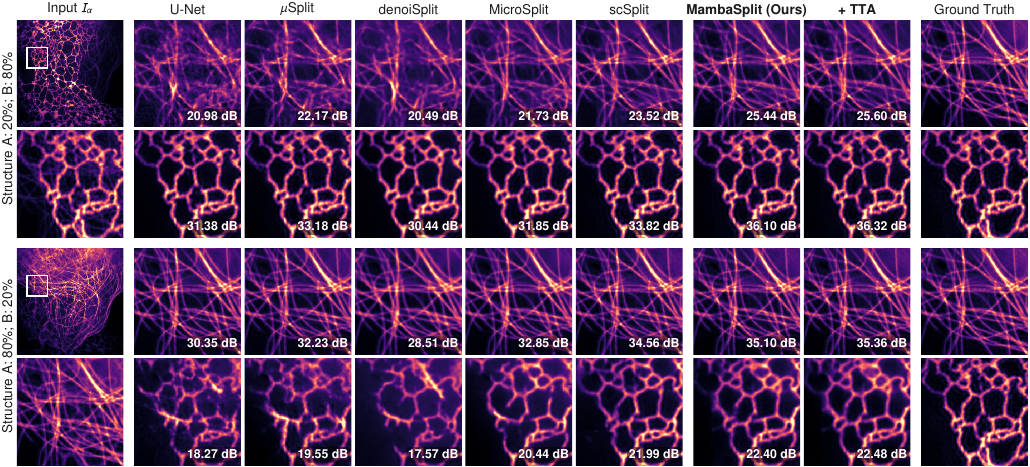}
\caption{\textbf{Weak- and dominant-component recovery on BioSR.} One test frame and crop at two mixing ratios: structure~A at $20\%$ and B at $80\%$ (top pair), and the reverse (bottom pair). Within each pair, the first row shows structure~A and the second structure~B. The input column shows the full frame with the crop marked, and the zoomed crop. Values report crop RI-PSNR (dB); $+$\,TTA is \MambaSplit{} with eight-view test-time augmentation. All baselines are the released models.}
\label{fig:gallery-BioSR}
\end{figure*}
\begin{figure*}[tbp]\centering
\includegraphics[width=\textwidth]{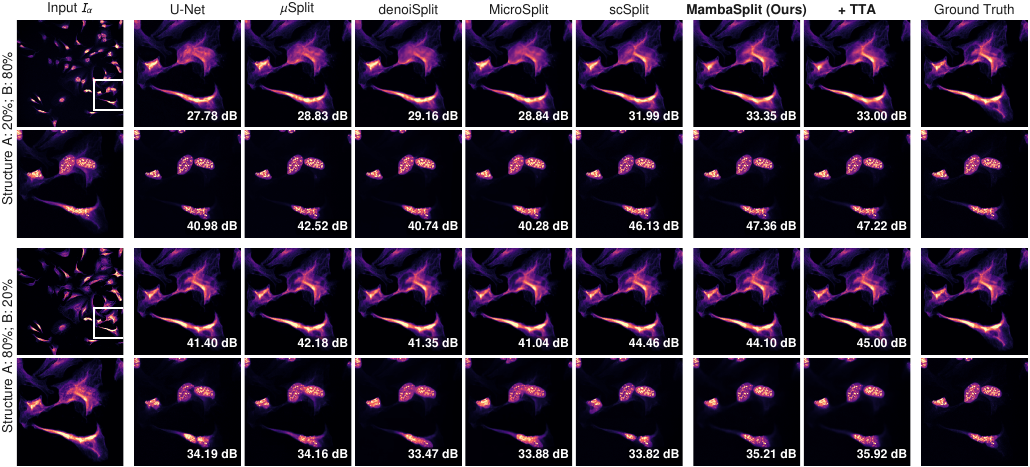}
\caption{\textbf{Weak- and dominant-component recovery on HT-LIF24.} One test frame and crop at two mixing ratios: structure~A at $20\%$ and B at $80\%$ (top pair), and the reverse (bottom pair). Within each pair, the first row shows structure~A and the second structure~B. The input column shows the full frame with the crop marked, and the zoomed crop. Values report crop RI-PSNR (dB); $+$\,TTA is \MambaSplit{} with eight-view test-time augmentation. All baselines are the released models.}
\label{fig:gallery-HTLIF24}
\end{figure*}
\begin{figure*}[tbp]\centering
\includegraphics[width=\textwidth]{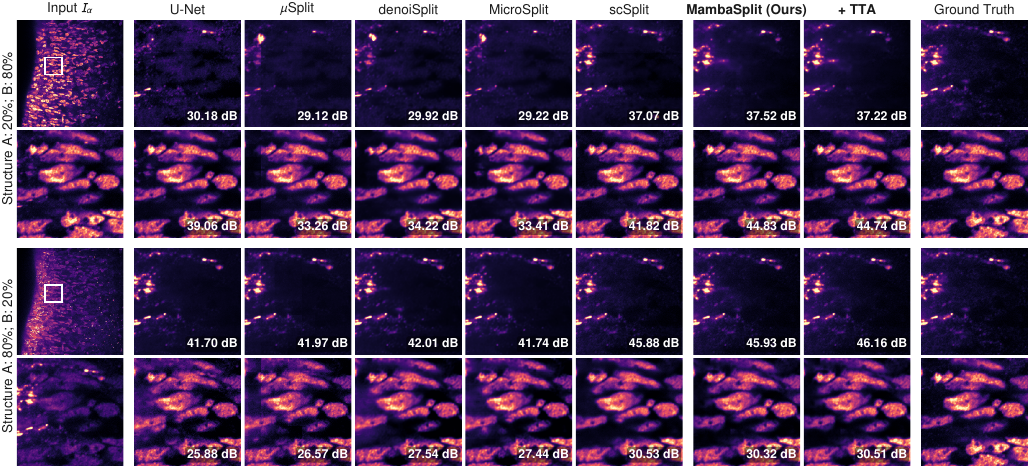}
\caption{\textbf{Weak- and dominant-component recovery on HT-T24.} One test frame and crop at two mixing ratios: structure~A at $20\%$ and B at $80\%$ (top pair), and the reverse (bottom pair). Within each pair, the first row shows structure~A and the second structure~B. The input column shows the full frame with the crop marked, and the zoomed crop. Values report crop RI-PSNR (dB); $+$\,TTA is \MambaSplit{} with eight-view test-time augmentation. All baselines are the released models.}
\label{fig:gallery-HTT24}
\end{figure*}
}

\newcommand{\figAcqQualMore}{%
\begin{figure*}[p]\centering
\includegraphics[width=0.75\textwidth]{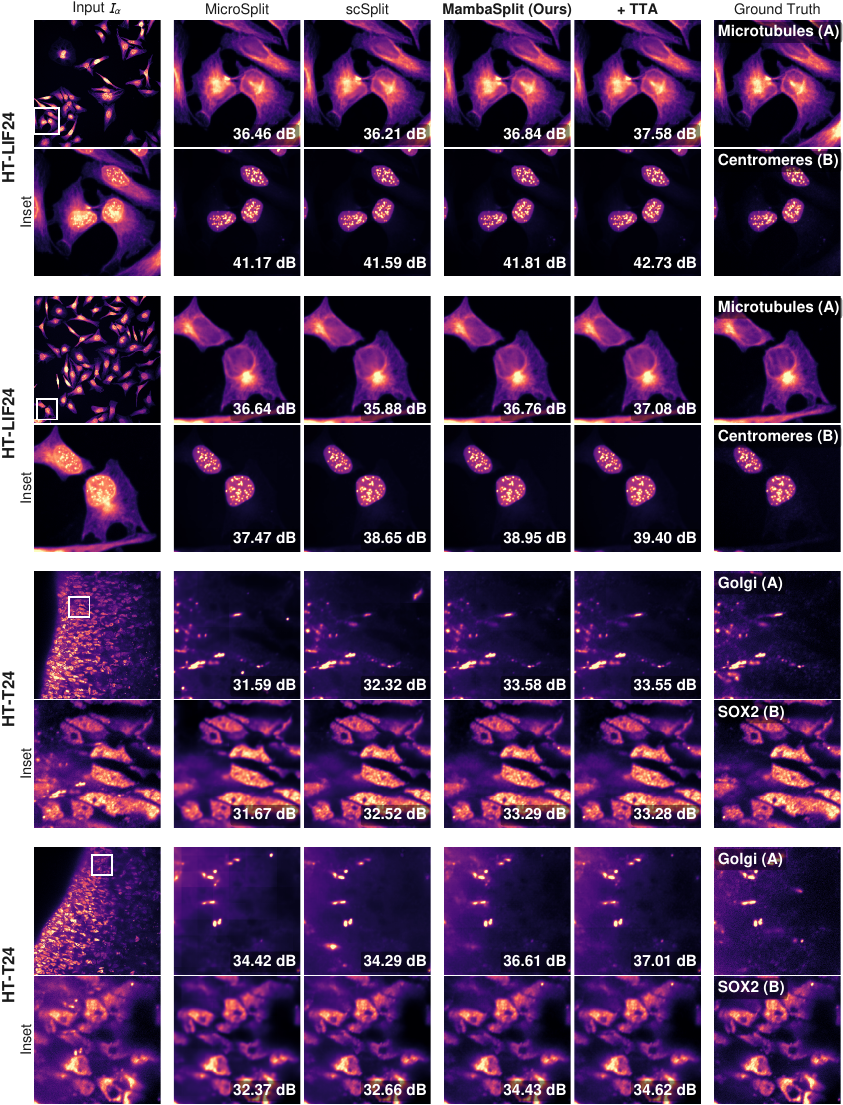}
\caption{\textbf{Additional reconstructions of experimentally acquired superpositions.} Two
additional test examples each from HT-LIF24 (microtubules, centromeres) and HT-T24 (Golgi, SOX2) after fine-tuning, complementing
\cref{fig:acquired-qual}. Crops are shared across methods and values report crop RI-PSNR (dB); $+$\,TTA is \MambaSplit{} with eight-view test-time augmentation. Models
and inference settings follow \supref{sec:supp:real}.}
\label{fig:acquired-qual-more}
\end{figure*}
}

\newcommand{\figRatioEstimationError}{%
\begin{figure}[t]
\centering
\includegraphics[width=0.92\columnwidth]{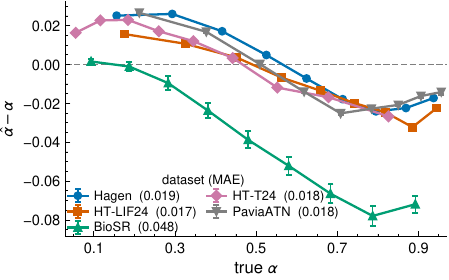}
\caption{\textbf{Mixing-ratio estimation error.} Mean residual $\hat{\alpha}-\alpha$ across test
tiles at each evaluated ratio; error bars show the standard error of the mean. Legend values give
the mean absolute error over all tiles. The dashed line denotes zero bias.}
\label{fig:ratio-estimation-error}
\end{figure}
}

\newcommand{\figSeamVsWeight}{%
\begin{figure*}[!h]\centering
\includegraphics[width=\textwidth]{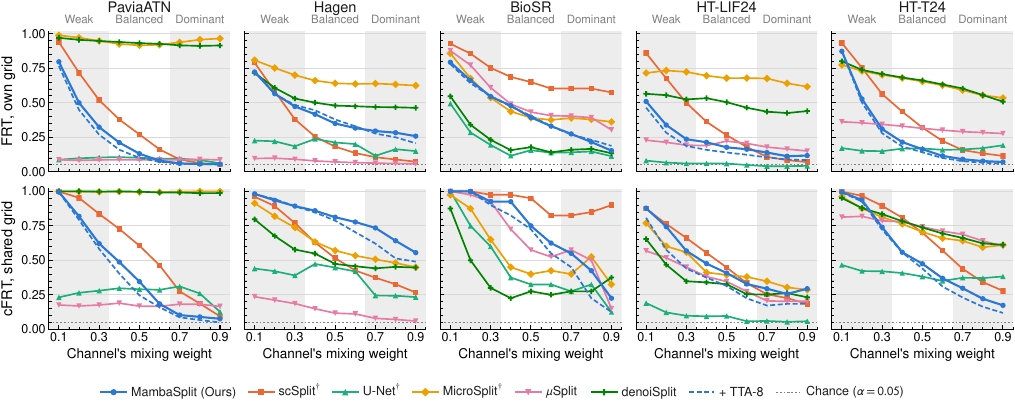}
\caption{\textbf{Tiling artifacts across mixing weights.}
FRT on each method's own grid (top) and cFRT on the shared $256$-pixel grid (bottom), against each output channel's mixing weight (channel~A has weight $w$, channel~B has $1-w$).
Shaded columns mark the Weak ($\le 0.3$) and Dominant ($\ge 0.7$) regimes; the dotted line is the nominal test level of $0.05$.
$^{\dagger}$Our unclipped retraining on PaviaATN and Hagen; other datasets use released models.}
\label{fig:seam-vs-weight}
\end{figure*}
}

\newcommand{\figArchMain}{%
\begin{figure*}[!htb]\centering
\includegraphics[width=\textwidth]{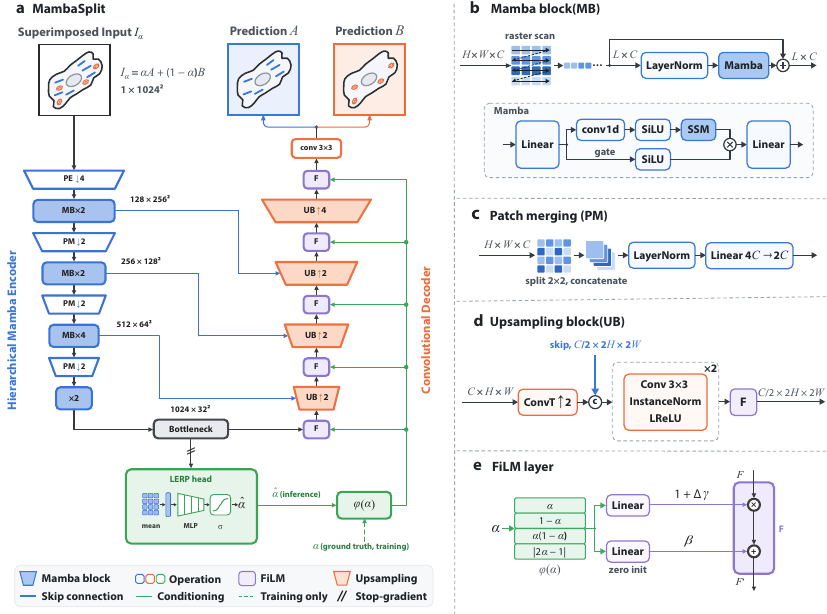}
\caption{\textbf{Detailed \MambaSplit{} architecture.} A $4\!\times\!4$ patch embedding and
hierarchical Mamba encoder progressively reduce spatial resolution while increasing channel width.
The LERP head predicts the mixing ratio from the bottleneck, and FiLM conditions the bottleneck and
shared convolutional decoder. Skip connections preserve spatial detail, and both output channels
are reconstructed jointly. Insets show the Mamba block, patch merging, decoder upsampling block and
FiLM operation; dimensions correspond to a $1024\!\times\!1024$ input.}
\label{fig:arch-main}
\end{figure*}
}

\newcommand{\figSeamQuality}{%
\begin{figure}[!tp]\centering
\includegraphics[width=0.87\columnwidth]{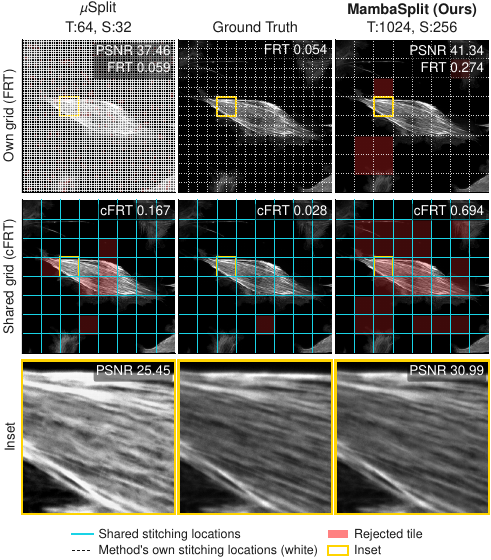}
\caption{\textbf{The FRT and cFRT scores alone are not a measure of reconstruction quality.}
\muSplit and \MambaSplit on one Hagen test frame ($w=0.5$), with the ground truth in the middle column.
$T$ and $S$ denote tile size and stride in pixels.
Top: FRT on each method's own grid; middle: cFRT on the shared $256$-pixel grid; red marks rejected tiles.
Bottom: the inset (yellow box) at higher magnification.
}
\label{fig:seam-quality}
\end{figure}
}

\newcommand{\figTilingArtefacts}{%
\begin{figure*}[!h]
\centering
\includegraphics[width=0.9\textwidth]{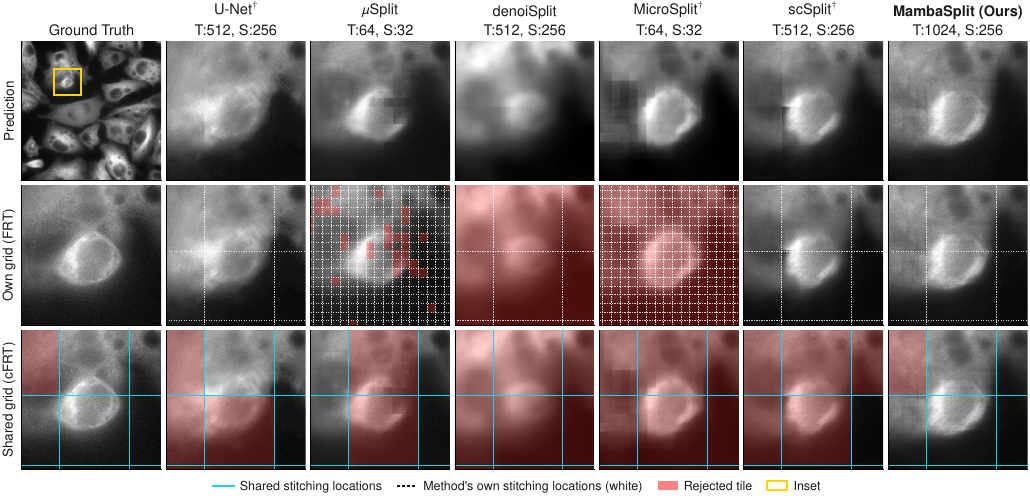}
\caption{\textbf{Tiling artifacts and their detection by FRT and cFRT.}
A $512$-pixel crop of a PaviaATN test frame (channel~B, $w=0.5$; yellow inset in the ground-truth frame).
$T$ and $S$ denote the tile size and stride, respectively, in pixels.
Top: predictions, where tiling artifacts appear as discontinuities at the method's tile boundaries.
Middle: FRT on each method's own grid (white dotted lines).
Bottom: cFRT on the shared $256$-pixel grid (cyan lines).
Red marks rejected tiles; on the $256$-pixel own grids, a tile is red when the median $p$-value over its $32$-pixel segment tests is below $0.05$.
}
\label{fig:tiling-artefacts}
\end{figure*}
}

\subsection*{S1\quad Implementation and evaluation details}\suplabel{sec:supp:impl}{S1}

\subsubsection*{S1.1\quad Datasets and training}
\suplabel{sec:supp:training}{S1.1}

\Cref{tab:datasets,tab:training-config} give the dataset splits and the default \MambaSplit{}
training configuration. Models trained on summed inputs use the final checkpoint unless stated
otherwise. Models trained from scratch on experimentally acquired superpositions use the
lowest-validation-loss checkpoint, whereas fine-tuned models are selected by validation RI-PSNR
(\supref{sec:supp:real}).

\begin{table}[!htb]\centering\footnotesize\setlength{\tabcolsep}{2.5pt}
\caption{\textbf{Dataset composition.} Unmixing task (channels A / B), image size, train/validation/test split, and number of
experimentally acquired superpositions in the test set. Splits follow scSplit~\cite{scsplit}. HT-LIF24 uses the $500$\,ms exposure.}
\label{tab:datasets}
\begin{tabular}{@{}l l r l r@{}}
\toprule
\textbf{Dataset} & \textbf{Structures (A / B)} & \textbf{Size} & \makecell{\textbf{Split}\\\textbf{(tr/val/test)}} & \makecell{\textbf{Acquired}\\\textbf{test inputs}} \\
\midrule
PaviaATN & Actin / Tubulin & $2720^2$ & 49/6/6    & / \\
Hagen    & Actin / Mitochondria & $2048^2$ & 80/10/10  & / \\
BioSR    & Microtubules / ER & $1004^2$ & 45/5/5    & / \\
\midrule
HT-LIF24 & Microtubules / Centromeres & $1608^2$ & 90/12/10  & 10 \\
HT-T24   & Golgi / SOX2 & $1608^2$ & 292/37/36 & 36 \\
\bottomrule
\end{tabular}
\end{table}

\begin{table}[!htb]
\centering\footnotesize\setlength{\tabcolsep}{4pt}
\caption{\textbf{Training configuration.} Default settings for
\MambaSplit{}.}
\label{tab:training-config}
\begin{tabular}{@{}l>{\raggedright\arraybackslash}p{0.62\columnwidth}@{}}
\toprule
Setting & Value \\
\midrule
\multicolumn{2}{@{}l}{\emph{Optimization}} \\
Optimizer & AdamW, lr $10^{-4}$, weight decay $10^{-2}$ \\
Schedule & cosine to $10^{-7}$, $299$ epochs, no warm-up \\
Steps & $2{,}000$ per epoch (batch $5$, i.e.\ $10{,}000$ random training crops per epoch) \\
Batch / patch & $5$ / $1024^2$ ($512^2$ on BioSR) \\
Precision & bf16 mixed \\
Seed / GPU & $955$ / single $35$\,GB \\
\midrule
\multicolumn{2}{@{}l}{\emph{Data}} \\
Ratio sampling & $\alpha\sim\mathrm{U}[0,1]$ \\
Normalization & fixed per-channel $z$-score \\
Augmentation & random rotation and reflection \\
Crops & frames drawn independently per channel (\mbox{\cref{tab:pairing}}); one shared crop position; no added noise or scaling \\
\midrule
\multicolumn{2}{@{}l}{\emph{Model and loss}} \\
Embedding / depths & $128$ / $[2,2,4,2]$ \\
State / conv / expand & $16$ / $4$ / $2$ \\
Decoder block & transposed conv $\uparrow2$, concat skip, $2\times$[conv $3\!\times\!3$, InstanceNorm, LeakyReLU] \\
Loss & Huber, $\delta=1.0$; ratio weight $\lambda=0.1$ \\
\bottomrule
\end{tabular}
\end{table}

For the patch-size ablation in \cref{fig:patch-size}, we
keep the number of training pixels per step fixed by adjusting
the batch size: $5$, $20$, and $80$ for $1024$-, $512$-, and
$256$-pixel patches, respectively. BioSR uses $512$- and
$256$-pixel patches with batches of $5$ and $20$.
All configurations use $2000$ steps per epoch. The default
models use their final checkpoints, while the smaller-patch
models use the best-validation checkpoints.
\Cref{tab:patch-size-pixmatch} reports their absolute RI-PSNR.

\begin{table}[!htb]
\centering
\small
\setlength{\tabcolsep}{6pt}

\caption{\textbf{Pixel-matched patch-size ablation.}
Nine-ratio mean RI-PSNR (dB), evaluated in a single pass
with inference stride equal to one quarter of the patch size.}
\label{tab:patch-size-pixmatch}

\begin{tabular}{@{}lccc@{}}
\toprule
& \multicolumn{3}{c}{\textbf{Training patch size}} \\
\cmidrule(lr){2-4}
\textbf{Dataset} & $1024$ & $512$ & $256$ \\
\midrule
PaviaATN & 32.81 & 32.49 & 31.62 \\
Hagen    & 41.33 & 41.14 & 40.83 \\
HT-LIF24 & 46.03 & 46.15 & 45.77 \\
HT-T24   & 40.45 & 40.16 & 39.26 \\
BioSR    &       & 37.31 & 35.34 \\
\bottomrule
\end{tabular}
\end{table}

\subsubsection*{S1.2\quad Architecture and conditioning}
\suplabel{sec:supp:archcond}{S1.2}

\cref{fig:arch-main} details the \MambaSplit{} architecture. A non-overlapping $4\!\times\!4$ patch
embedding produces width $128$, followed by four encoder stages with widths $128$, $256$, $512$
and $1024$ and depths $[2,2,4,2]$. Between stages, $2\!\times\!2$ patch merging~\cite{swin} halves
the spatial resolution and doubles the channel width. Each Mamba block~\cite{mamba} uses state,
convolution and expansion dimensions $(16,4,2)$ with a single forward row-major scan.

\figArchMain

The LERP head globally pools the bottleneck and predicts $\hat{\alpha}$ using a three-layer MLP
with hidden widths $128$ and $64$. FiLM conditioning~\cite{film} is applied at the bottleneck and
decoder stages using
\begin{equation}
  \varphi(\alpha) = \bigl[\,\alpha,\; 1-\alpha,\; \alpha(1-\alpha),\;
  \lvert 2\alpha-1 \rvert \,\bigr],
  \label{eq:phi}
\end{equation}
with learned channel-wise residual scale and bias,
\begin{equation}
  h'_l = \bigl[\,1 + \Delta\gamma_l\bigl(\varphi(\alpha)\bigr)\,\bigr]
  \odot h_l + \beta_l\bigl(\varphi(\alpha)\bigr).
  \label{eq:film}
\end{equation}
The terms $\alpha$ and $1-\alpha$ retain directional information about which component is dominant;
$\alpha(1-\alpha)$ is maximal for balanced mixtures and decreases symmetrically toward the extremes;
and $\lvert 2\alpha-1\rvert$ encodes the magnitude of imbalance independently of which component
dominates. We use this four-term map as a compact severity representation for FiLM conditioning; \cref{tab:ratio-conditioning-ablation} compares it with conditioning on $\hat{\alpha}$ alone.
The FiLM projections are initialized to zero. The shared decoder upsamples with skip connections
and predicts both output channels jointly.

\subsubsection*{S1.3\quad Inference and evaluation}
\suplabel{sec:supp:targets}{S1.3}\suplabel{sec:supp:axis}{S1.3}

\MambaSplit{} uses overlapping $1024^2$ tiles ($512^2$ for BioSR) at stride $256$ ($128$ for BioSR), one quarter of the tile size, so that every interior pixel is covered by $16$ tiles, and averages
overlapping predictions. The default evaluation uses one prediction per tile; TTA averages
predictions over the eight dihedral transformations before de-normalization.

In the released Hagen and PaviaATN data, both target channels are upper-clipped at a single
shared threshold: the $99.5$th percentile of the training-split pixels pooled over both channels
($1993$ for Hagen, $1308$ for PaviaATN), applied to all splits. We refer to the source targets
before this operation as \emph{unclipped} and use one target definition per dataset for all
methods: the unclipped source channels for Hagen, and, for PaviaATN, channel~A rebuilt from the raw
source data together with the released channel~B, whose test intensities stay below the threshold
and are therefore unaffected.
BioSR, HT-LIF24 and HT-T24 are scored as released. Predictions are not clipped.

scSplit parameterizes mixtures by the raw channel-A fraction $w$, whereas \MambaSplit{} is trained
using the standardized ratio $\alpha$. We therefore convert using
\begin{equation*}
  w(\alpha)=\frac{\alpha\sigma_B}{\alpha\sigma_B+(1-\alpha)\sigma_A}.
\end{equation*}
For scSplit-axis comparisons (\cref{tab:summed-severity-results} and \cref{tab:perratio-a,tab:perratio-b}), \MambaSplit{}
is evaluated at the corresponding $\alpha$.

\subsection*{S2\quad Reconstruction across mixing ratios}
\suplabel{sec:supp:perratio}{S2}

Because some published PaviaATN and Hagen baselines were trained or evaluated against clipped targets, we retrain the affected methods with their original recipes using the common unclipped targets defined in \supref{sec:supp:targets} and mark them $^{\dagger}$.

\subsubsection*{S2.1\quad Per-image comparison}\suplabel{sec:supp:perimage}{S2.1}

\cref{tab:per-image} breaks the regime means of \cref{tab:summed-severity-results} down to individual test images. In a single pass, \MambaSplit{} is better than the strongest baseline on $148$ of $201$ image--regime pairs ($74\%$), within $0.1$\,dB on $22$ and worse on $31$, most of them on Hagen; with TTA it is better on $176$ ($88\%$) and worse on $12$.

\begin{table*}[!tp]
\centering
\small
\setlength{\tabcolsep}{5pt}

\caption{\textbf{Per-image comparison with the strongest baseline.} For each test image and regime, the mean RI-PSNR over the regime's six channel--ratio cells. $\Delta$ is \MambaSplit{} minus the strongest baseline of \cref{tab:summed-severity-results}, which is scSplit in every cell (scSplit$^{\dagger}$ on PaviaATN and Hagen), reported as mean $\pm$ s.d.\ over images (dB). W, D and L give the percentage of images won, drawn ($|\Delta|<0.1$\,dB), and lost. Dominant, Balanced and Weak as in  \cref{tab:summed-severity-results}.}
\label{tab:per-image}

\begin{tabular}{@{}ll cccc cccc@{}}
\toprule
& & \multicolumn{4}{c}{\textbf{Single pass}} & \multicolumn{4}{c}{\textbf{+TTA}} \\
\cmidrule(lr){3-6}\cmidrule(lr){7-10}
\textbf{Dataset} & \textbf{Regime} & $\Delta$ & W & D & L & $\Delta$ & W & D & L \\
\midrule

\multirow{3}{*}{PaviaATN}
& Dominant  & $+0.78\pm0.15$ & 100\% & 0\%  & 0\%   & $+1.31\pm0.12$ & 100\% & 0\%  & 0\% \\
& Balanced  & $-0.21\pm0.15$ & 0\%   & 17\% & 83\%  & $+0.40\pm0.22$ & 100\% & 0\%  & 0\% \\
& Weak & $+0.80\pm0.27$ & 100\% & 0\%  & 0\%   & $+1.15\pm0.32$ & 100\% & 0\%  & 0\% \\
\midrule

\multirow{3}{*}{Hagen}
& Dominant  & $-0.83\pm0.45$ & 10\%  & 0\%  & 90\%  & $-0.28\pm0.40$ & 20\%  & 0\%  & 80\% \\
& Balanced  & $-0.37\pm0.29$ & 10\%  & 10\% & 80\%  & $+0.03\pm0.43$ & 40\%  & 20\% & 40\% \\
& Weak & $+0.25\pm0.33$ & 50\%  & 50\% & 0\%   & $+0.53\pm0.37$ & 90\%  & 10\% & 0\% \\
\midrule

\multirow{3}{*}{BioSR}
& Dominant  & $+3.01\pm1.32$ & 100\% & 0\% & 0\% & $+3.71\pm1.39$ & 100\% & 0\% & 0\% \\
& Balanced  & $+2.39\pm0.91$ & 100\% & 0\% & 0\% & $+2.83\pm1.17$ & 100\% & 0\% & 0\% \\
& Weak & $+2.40\pm0.50$ & 100\% & 0\% & 0\% & $+2.85\pm0.55$ & 100\% & 0\% & 0\% \\
\midrule

\multirow{3}{*}{HT-LIF24}
& Dominant  & $+0.45\pm0.36$ & 70\% & 30\% & 0\%  & $+1.08\pm0.34$ & 100\% & 0\%  & 0\% \\
& Balanced  & $+0.09\pm0.38$ & 60\% & 10\% & 30\% & $+0.53\pm0.40$ & 80\%  & 20\% & 0\% \\
& Weak & $+0.36\pm0.31$ & 80\% & 20\% & 0\%  & $+0.72\pm0.30$ & 100\% & 0\%  & 0\% \\
\midrule

\multirow{3}{*}{HT-T24}
& Dominant  & $+1.89\pm0.64$ & 100\% & 0\%  & 0\%  & $+2.62\pm0.84$ & 100\% & 0\%  & 0\% \\
& Balanced  & $+0.68\pm0.47$ & 86\%  & 14\% & 0\%  & $+0.86\pm0.63$ & 94\%  & 6\%  & 0\% \\
& Weak & $+0.38\pm0.51$ & 72\%  & 11\% & 17\% & $+0.59\pm0.60$ & 83\%  & 17\% & 0\% \\
\midrule

\multicolumn{2}{c}{Across datasets}
& $+0.76\pm1.02$ & 74\% & 11\% & 15\%
& $+1.18\pm1.17$ & 88\% & 6\% & 6\% \\
\bottomrule
\end{tabular}
\end{table*}

\subsubsection*{S2.2\quad Additional metrics}\suplabel{sec:supp:extended-metrics}{S2.2}

For the additional metrics in \cref{tab:extended-metrics}, we compute LPIPS~\cite{lpips} using AlexNet after scaling prediction and target to $[-1,1]$ using the target intensity range, and PSNR in raw intensity units using the target range as the data range. We use released prediction stacks when they reproduce the published RI-PSNR and our own predictions otherwise. We do not report published LPIPS values because their implementation does not reproduce consistently on HT-LIF24 and HT-T24. Unlike RI-PSNR, PSNR does not remove global affine intensity differences, which accounts for the large discrepancy between the two metrics for several released baselines.

\begin{table*}[!tp]
\centering
\caption{\textbf{Extended metrics across mixing severities.}
RI-PSNR, MS-SSIM, LPIPS, and PSNR in the Dominant (D), Balanced (B), and Weak (W) regimes for the methods of \cref{tab:summed-severity-results}. Channels are grouped by their own mixture weight: D $=\{0.7,0.8,0.9\}$, B $=\{0.4,0.5,0.6\}$, and W $=\{0.1,0.2,0.3\}$; values average both channels, the three corresponding weights, and all test frames. RI-PSNR and MS-SSIM reproduce \cref{tab:summed-severity-results}. Rows marked $^{\dagger}$ are our retraining on raw, unclipped data; published PaviaATN and Hagen rows use their original clipped-target protocol and are excluded from ranking. LPIPS and PSNR are computed consistently from available prediction stacks. Dashes indicate unavailable predictions; TTA denotes test-time augmentation. Best and second-best comparable results per dataset and metric are \best{bold} and \nd{underlined}.}
\label{tab:extended-metrics}
\footnotesize\setlength{\tabcolsep}{3.5pt}\renewcommand{\arraystretch}{0.79}
\begin{adjustbox}{max width=\textwidth}
\begin{tabular}{@{}c l ccc@{\hspace{1.2em}}ccc@{\hspace{1.2em}}ccc@{\hspace{1.2em}}ccc@{}}
\toprule
\multirow{2}{*}{\textbf{Data}} & \multirow{2}{*}{\textbf{Model}} & \multicolumn{3}{c}{\textbf{RI-PSNR (dB)} $\,\uparrow$} & \multicolumn{3}{c}{\textbf{MS-SSIM} $\,\uparrow$} & \multicolumn{3}{c}{\textbf{LPIPS} $\,\downarrow$} & \multicolumn{3}{c}{\textbf{PSNR (dB)} $\,\uparrow$} \\
\cmidrule(lr){3-5}\cmidrule(lr){6-8}\cmidrule(lr){9-11}\cmidrule(l){12-14}
 & & D & B & W & D & B & W & D & B & W & D & B & W \\
\midrule
\multirow{11}{*}{\rotatebox[origin=c]{90}{\textbf{PaviaATN}}}
 & U-Net & 29.3 & 25.4 & 21.2 & 0.870 & 0.743 & 0.568 & 0.222 & 0.365 & 0.519 & 15.8 & 11.7 & 8.2 \\
 & $\mu$Split & 27.9 & 25.2 & 21.3 & 0.908 & 0.808 & 0.648 & 0.136 & 0.256 & 0.416 & 21.3 & 15.2 & 10.3 \\
 & denoiSplit & 27.3 & 26.2 & 21.8 & 0.857 & 0.843 & 0.750 & 0.806 & 0.830 & 0.862 & 22.7 & 18.4 & 13.6 \\
 & MicroSplit & 24.0 & 21.8 & 18.8 & 0.771 & 0.688 & 0.540 & 0.910 & 0.960 & 1.004 & 14.7 & 10.1 & 6.3 \\
 & InDI & 29.9 & 23.9 & 21.7 & 0.943 & 0.858 & 0.741 & 0.134 & 0.195 & 0.251 & 24.5 & 21.7 & 20.5 \\
 & scSplit & 35.1 & 27.6 & 24.3 & 0.977 & 0.907 & 0.823 & 0.035 & 0.158 & 0.382 & 33.3 & 27.0 & 24.1 \\
 \cmidrule(l){2-14}
 & U-Net$^{\dagger}$ & 30.7 & 26.9 & 21.8 & 0.902 & 0.816 & 0.647 & 0.137 & 0.253 & 0.421 & 19.4 & 14.9 & 10.5 \\
 & MicroSplit$^{\dagger}$ & 26.1 & 26.7 & 22.3 & 0.862 & 0.867 & 0.763 & 0.815 & 0.806 & 0.861 & 22.6 & 25.8 & 17.3 \\
 & scSplit$^{\dagger}$ & 39.4 & \nd{30.4} & 27.3 & 0.980 & \nd{0.918} & \nd{0.862} & 0.031 & 0.143 & 0.437 & 39.1 & \nd{30.4} & 27.1 \\
\cmidrule(l){3-14}
 & \textbf{\MambaSplit{}} & \nd{40.1} & 30.2 & \nd{28.1} & \nd{0.981} & 0.915 & 0.859 & \nd{0.030} & \nd{0.138} & \nd{0.406} & \nd{40.0} & 30.2 & \nd{27.8} \\
 & \quad $+$ TTA & \best{40.7} & \best{30.8} & \best{28.4} & \best{0.982} & \best{0.921} & \best{0.865} & \best{0.029} & \best{0.133} & \best{0.402} & \best{40.5} & \best{30.8} & \best{28.1} \\
\midrule
\multirow{11}{*}{\rotatebox[origin=c]{90}{\textbf{Hagen}}}
 & U-Net & 31.8 & 28.2 & 22.0 & 0.965 & 0.921 & 0.833 & -- & -- & -- & -- & -- & -- \\
 & $\mu$Split & 33.9 & 32.4 & 23.6 & 0.962 & 0.960 & 0.858 & 0.046 & 0.062 & 0.166 & 24.6 & 26.8 & 21.3 \\
 & denoiSplit & 32.4 & 31.9 & 23.1 & 0.958 & 0.954 & 0.851 & 0.165 & 0.169 & 0.246 & 24.7 & 26.0 & 20.8 \\
 & MicroSplit & 34.7 & 31.7 & 25.0 & 0.957 & 0.956 & 0.862 & 0.166 & 0.168 & 0.240 & 23.6 & 26.2 & 21.6 \\
 & InDI & 33.1 & 32.1 & 24.2 & 0.963 & 0.965 & 0.879 & 0.043 & 0.052 & 0.138 & 26.0 & 29.7 & 23.0 \\
 & scSplit & 40.9 & 33.9 & 29.3 & 0.994 & 0.977 & 0.934 & 0.010 & 0.046 & 0.122 & 40.0 & 33.0 & 28.1 \\
 \cmidrule(l){2-14}
 & U-Net$^{\dagger}$ & 43.8 & 39.5 & 31.6 & 0.971 & 0.976 & 0.914 & 0.035 & 0.039 & 0.136 & 29.9 & 33.8 & 28.2 \\
 & MicroSplit$^{\dagger}$ & 40.7 & 38.1 & 31.8 & \nd{0.974} & 0.980 & 0.921 & 0.103 & 0.099 & 0.168 & 30.3 & 36.6 & 28.7 \\
 & scSplit$^{\dagger}$ & \best{48.0} & \best{40.9} & 36.0 & \best{0.997} & \best{0.988} & \best{0.964} & \best{0.006} & \best{0.026} & \best{0.076} & 45.7 & 40.2 & \nd{35.7} \\
\cmidrule(l){3-14}
 & \textbf{\MambaSplit{}} & 47.2 & \nd{40.5} & \nd{36.3} & \best{0.997} & \nd{0.987} & \nd{0.962} & \nd{0.008} & \nd{0.029} & \nd{0.080} & \nd{46.5} & \nd{40.4} & 35.6 \\
 & \quad $+$ TTA & \nd{47.7} & \best{40.9} & \best{36.6} & \best{0.997} & \best{0.988} & \best{0.964} & \nd{0.008} & \nd{0.029} & 0.083 & \best{47.1} & \best{40.8} & \best{35.9} \\
\midrule
\multirow{8}{*}{\rotatebox[origin=c]{90}{\textbf{BioSR}}}
 & U-Net & 37.2 & 33.7 & 25.6 & 0.924 & 0.958 & 0.740 & 0.066 & 0.059 & 0.291 & 24.2 & 31.1 & 22.2 \\
 & $\mu$Split & 37.8 & 33.5 & 25.7 & 0.918 & 0.959 & 0.738 & 0.065 & 0.051 & 0.291 & 23.3 & 30.9 & 21.8 \\
 & denoiSplit & 36.4 & 33.1 & 25.3 & 0.929 & 0.957 & 0.733 & 0.083 & 0.086 & 0.322 & 24.3 & 30.8 & 21.9 \\
 & MicroSplit & 38.5 & 34.3 & 26.6 & 0.932 & 0.966 & 0.759 & 0.068 & 0.065 & 0.274 & 24.1 & 31.1 & 22.2 \\
 & InDI & 35.9 & 33.4 & 26.3 & 0.917 & 0.953 & 0.802 & 0.054 & 0.051 & 0.211 & 24.2 & 29.3 & 23.8 \\
 & scSplit & 40.1 & 35.3 & 28.7 & 0.987 & 0.973 & 0.889 & 0.011 & 0.032 & 0.130 & 36.8 & 32.8 & 26.8 \\
\cmidrule(l){3-14}
 & \textbf{\MambaSplit{}} & \nd{43.1} & \nd{37.7} & \nd{31.1} & \nd{0.993} & \nd{0.982} & \nd{0.922} & \nd{0.009} & \nd{0.025} & \nd{0.096} & \nd{39.7} & \best{34.7} & \nd{28.8} \\
 & \quad $+$ TTA & \best{43.9} & \best{38.1} & \best{31.5} & \best{0.994} & \best{0.983} & \best{0.926} & \best{0.008} & \best{0.024} & \best{0.093} & \best{39.9} & \nd{34.6} & \best{28.9} \\
\midrule
\multirow{8}{*}{\rotatebox[origin=c]{90}{\textbf{HT-LIF24}}}
 & U-Net & 45.9 & 44.6 & 36.0 & 0.980 & 0.986 & 0.939 & 0.025 & 0.020 & 0.091 & 32.8 & 37.0 & 32.4 \\
 & $\mu$Split & 46.7 & 45.1 & 36.6 & 0.978 & 0.986 & 0.940 & 0.026 & 0.020 & 0.091 & 32.4 & 36.8 & 32.4 \\
 & denoiSplit & 44.8 & 42.9 & 35.8 & 0.981 & 0.985 & 0.938 & 0.046 & 0.044 & 0.108 & 33.1 & 36.9 & 32.0 \\
 & MicroSplit & 45.0 & 43.7 & 36.5 & 0.982 & 0.986 & 0.939 & 0.047 & 0.045 & 0.106 & 33.5 & 37.2 & 32.2 \\
 & InDI & 45.2 & 43.9 & 37.6 & 0.976 & 0.991 & 0.963 & 0.032 & 0.018 & 0.071 & 32.1 & 39.6 & 34.0 \\
 & scSplit & 51.8 & \nd{45.5} & 39.9 & \nd{0.998} & \nd{0.994} & \nd{0.976} & \nd{0.004} & \nd{0.014} & 0.048 & 47.1 & 40.7 & 35.9 \\
\cmidrule(l){3-14}
 & \textbf{\MambaSplit{}} & \nd{52.3} & \nd{45.5} & \nd{40.3} & \best{0.999} & \best{0.996} & \best{0.988} & \best{0.003} & \best{0.011} & \best{0.034} & \nd{51.5} & \nd{44.9} & \nd{39.3} \\
 & \quad $+$ TTA & \best{52.9} & \best{46.0} & \best{40.6} & \best{0.999} & \best{0.996} & \best{0.988} & \best{0.003} & \best{0.011} & \nd{0.035} & \best{52.0} & \best{45.1} & \best{39.5} \\
\midrule
\multirow{8}{*}{\rotatebox[origin=c]{90}{\textbf{HT-T24}}}
 & U-Net & 37.9 & 37.5 & 30.1 & 0.963 & 0.965 & 0.883 & 0.097 & 0.095 & 0.199 & 30.3 & 29.9 & 26.6 \\
 & $\mu$Split & 37.6 & 36.9 & 29.9 & 0.954 & 0.959 & 0.880 & 0.106 & 0.097 & 0.205 & 29.7 & 30.1 & 26.4 \\
 & denoiSplit & 37.3 & 37.5 & 31.0 & 0.954 & 0.964 & 0.896 & 0.339 & 0.314 & 0.362 & 29.0 & 30.6 & 27.0 \\
 & MicroSplit & 36.9 & 36.6 & 30.3 & 0.950 & 0.959 & 0.891 & 0.351 & 0.324 & 0.369 & 29.0 & 29.5 & 26.4 \\
 & InDI & 37.6 & 36.5 & 30.5 & 0.962 & 0.966 & 0.909 & 0.095 & 0.085 & \best{0.166} & 30.7 & 33.2 & 29.6 \\
 & scSplit & 44.5 & 39.1 & 34.7 & \nd{0.995} & \nd{0.981} & \nd{0.943} & \nd{0.018} & \best{0.067} & \nd{0.192} & 42.3 & 38.0 & 33.4 \\
\cmidrule(l){3-14}
 & \textbf{\MambaSplit{}} & \nd{46.5} & \nd{39.8} & \nd{35.1} & \nd{0.995} & 0.980 & 0.942 & \nd{0.018} & \best{0.067} & 0.199 & \nd{46.0} & \nd{39.3} & \nd{34.7} \\
 & \quad $+$ TTA & \best{47.2} & \best{40.0} & \best{35.3} & \best{0.996} & \best{0.982} & \best{0.944} & \best{0.017} & \nd{0.068} & 0.206 & \best{46.7} & \best{39.5} & \best{34.8} \\
\bottomrule
\end{tabular}
\end{adjustbox}
\end{table*}

\subsubsection*{S2.3\quad Full per-ratio values}\suplabel{sec:supp:perratio-values}{S2.3}

\cref{tab:perratio-a,tab:perratio-b} list the RI-PSNR of both channels at every released ratio, for
every method with per-ratio predictions; their regime means reproduce \cref{tab:summed-severity-results}.
Across datasets, \MambaSplit{}'s advantage is generally largest when the target component is weak and smaller in Balanced and Dominant mixtures. On PaviaATN channel~A, for example, scSplit$^{\dagger}$ is $0.56$\,dB higher on average over $w=0.5$--$0.9$, whereas \MambaSplit{} is $1.26$\,dB higher over $w=0.1$--$0.3$ (\cref{tab:perratio-a}).

\begin{table*}[p]\centering
\caption{\textbf{RI-PSNR at every mixing ratio: PaviaATN and Hagen.} RI-PSNR (dB) of channels A and
B at the nine released ratios $w$. The Dominant, Balanced and Weak means of
\cref{tab:summed-severity-results} average each channel's own weight over $\{0.7,0.8,0.9\}$,
$\{0.4,0.5,0.6\}$ and $\{0.1,0.2,0.3\}$. $^{\dagger}$Our retraining on unclipped data.
Best and second-best among the
unclipped-target rows, including TTA, are \best{bold} and \nd{underlined}.}
\label{tab:perratio-a}
\footnotesize\setlength{\tabcolsep}{4.5pt}\renewcommand{\arraystretch}{0.92}
\begin{adjustbox}{max width=\textwidth}
\begin{tabular}{@{}c c l ccccccccc@{}}
\toprule
& & & \multicolumn{9}{c}{Channel-A weight $w$} \\
\cmidrule(l){4-12}
\textbf{Dataset} & \textbf{Ch.} & \textbf{Model} & 0.1 & 0.2 & 0.3 & 0.4 & 0.5 & 0.6 & 0.7 & 0.8 & 0.9 \\
\midrule
 \multirow{22}{*}{\rotatebox[origin=c]{90}{\textbf{PaviaATN}}} & \multirow{11}{*}{A} & U-Net & 20.90 & 23.25 & 25.82 & 27.94 & 29.65 & 30.94 & 31.96 & 32.76 & 33.34 \\
  &  & U-Net$^{\dagger}$ & 20.95 & 23.41 & 26.27 & 29.13 & 31.85 & 34.01 & 35.12 & 35.13 & 34.51 \\
  &  & $\mu$Split & 21.56 & 23.64 & 26.00 & 28.09 & 29.64 & 30.48 & 30.82 & 30.91 & 30.86 \\
  &  & denoiSplit & 21.08 & 22.98 & 25.47 & 28.14 & 30.37 & 31.69 & 32.22 & 32.39 & 32.38 \\
  &  & MicroSplit & 16.20 & 18.82 & 21.21 & 22.23 & 22.93 & 23.50 & 23.81 & 23.99 & 24.13 \\
  &  & MicroSplit$^{\dagger}$ & 21.17 & 24.29 & 26.92 & 29.03 & 30.07 & 29.94 & 29.13 & 28.22 & 27.71 \\
  &  & InDI & 21.15 & 25.52 & 26.84 & 25.72 & 25.29 & 25.68 & 26.79 & 28.13 & 29.17 \\
  &  & scSplit & 23.03 & 25.48 & 27.17 & 28.07 & 29.14 & 31.81 & 34.83 & 36.93 & 37.07 \\
  &  & scSplit$^{\dagger}$ & 29.44 & 31.39 & 32.68 & 33.62 & \best{34.93} & \nd{37.74} & 41.08 & \nd{44.80} & \nd{49.31} \\
  &  & \MambaSplit{} & \nd{30.58} & \nd{32.68} & \nd{34.02} & \nd{34.54} & 33.40 & 37.09 & \nd{41.20} & 44.77 & 48.58 \\
  &  & \quad $+$ TTA & \best{30.78} & \best{32.86} & \best{34.23} & \best{34.91} & \nd{34.14} & \best{37.92} & \best{41.86} & \best{45.46} & \best{49.73} \\
\cmidrule(l){2-12}
  & \multirow{11}{*}{B} & U-Net & 27.93 & 25.70 & 23.92 & 22.37 & 21.16 & 20.24 & 19.59 & 19.12 & 18.80 \\
  &  & U-Net$^{\dagger}$ & 28.07 & 26.37 & 24.73 & 23.21 & 22.04 & 21.17 & 20.52 & 20.04 & 19.69 \\
  &  & $\mu$Split & 25.62 & 25.22 & 24.00 & 22.31 & 20.91 & 19.91 & 19.24 & 18.79 & 18.49 \\
  &  & denoiSplit & 22.23 & 22.32 & 22.47 & 22.66 & 22.53 & 22.00 & 21.21 & 20.47 & 19.85 \\
  &  & MicroSplit & 25.00 & 24.02 & 23.34 & 21.77 & 20.60 & 19.86 & 19.30 & 18.89 & 18.61 \\
  &  & MicroSplit$^{\dagger}$ & 23.14 & 23.93 & 24.62 & 24.72 & 24.16 & 22.54 & 20.96 & 20.29 & 19.98 \\
  &  & InDI & 32.52 & 32.60 & 30.37 & 25.33 & 21.76 & 19.90 & 19.15 & 18.83 & 18.68 \\
  &  & scSplit & 38.13 & 33.46 & 30.18 & 26.95 & 24.95 & 24.52 & 24.10 & 23.45 & 22.61 \\
  &  & scSplit$^{\dagger}$ & 37.72 & 33.18 & 30.02 & 26.98 & \best{24.93} & \nd{24.41} & 24.00 & 23.47 & \nd{22.79} \\
  &  & \MambaSplit{} & \nd{39.63} & \nd{34.92} & \nd{31.68} & \nd{28.38} & 23.78 & 24.14 & \nd{24.65} & \nd{24.06} & 22.56 \\
  &  & \quad $+$ TTA & \best{39.90} & \best{35.12} & \best{31.88} & \best{28.76} & \nd{24.43} & \best{24.83} & \best{25.20} & \best{24.56} & \best{23.04} \\
\midrule
 \multirow{20}{*}{\rotatebox[origin=c]{90}{\textbf{Hagen}}} & \multirow{10}{*}{A} & U-Net$^{\dagger}$ & 27.02 & 31.03 & 34.10 & 36.44 & 38.09 & 39.18 & 39.86 & 40.90 & 43.87 \\
  &  & $\mu$Split & 20.11 & 23.52 & 27.46 & 31.12 & 33.00 & 32.68 & 32.52 & 32.84 & 33.55 \\
  &  & denoiSplit & 19.97 & 22.89 & 26.44 & 30.06 & 32.19 & 32.06 & 31.29 & 30.82 & 31.25 \\
  &  & MicroSplit & 22.12 & 25.83 & 28.54 & 30.41 & 31.67 & 32.53 & 33.25 & 33.99 & 35.56 \\
  &  & MicroSplit$^{\dagger}$ & 26.78 & 30.56 & 33.15 & 34.85 & 35.92 & 36.59 & 37.05 & 37.51 & 38.36 \\
  &  & InDI & 22.00 & 26.22 & 30.26 & 33.01 & 32.60 & 30.30 & 29.22 & 29.11 & 29.42 \\
  &  & scSplit & 28.62 & 31.35 & 32.75 & 33.26 & 33.54 & 34.03 & 35.47 & 37.89 & 42.49 \\
  &  & scSplit$^{\dagger}$ & 34.07 & 36.43 & 37.72 & \nd{38.46} & \best{39.26} & \nd{40.45} & \nd{42.20} & \best{44.62} & \best{48.69} \\
  &  & \MambaSplit{} & \nd{34.64} & \nd{36.89} & \nd{37.96} & 38.27 & 38.65 & 40.09 & 41.79 & 43.88 & 46.65 \\
  &  & \quad $+$ TTA & \best{34.88} & \best{37.18} & \best{38.28} & \best{38.60} & \nd{39.12} & \best{40.57} & \best{42.25} & \nd{44.41} & \nd{47.39} \\
\cmidrule(l){2-12}
  & \multirow{10}{*}{B} & U-Net$^{\dagger}$ & 46.72 & 46.29 & 45.09 & 43.42 & 41.30 & 38.72 & 35.62 & 32.15 & 29.74 \\
  &  & $\mu$Split & 34.52 & 34.76 & 35.01 & 34.77 & 33.16 & 29.76 & 26.15 & 23.11 & 21.17 \\
  &  & denoiSplit & 32.94 & 33.53 & 34.27 & 34.57 & 33.03 & 29.64 & 25.91 & 22.46 & 20.69 \\
  &  & MicroSplit & 35.43 & 35.15 & 34.53 & 33.50 & 32.00 & 30.00 & 27.55 & 24.16 & 21.98 \\
  &  & MicroSplit$^{\dagger}$ & 44.48 & 43.81 & 43.00 & 41.97 & 40.55 & 38.68 & 36.37 & 33.54 & 30.55 \\
  &  & InDI & 35.98 & 36.85 & 37.75 & 36.84 & 32.55 & 27.56 & 23.98 & 21.95 & 20.94 \\
  &  & scSplit & 46.73 & 42.98 & 40.00 & 36.90 & 34.18 & 31.78 & 29.87 & 27.71 & 25.51 \\
  &  & scSplit$^{\dagger}$ & \best{54.27} & \nd{50.66} & \nd{47.72} & \best{44.89} & \best{42.22} & \nd{40.08} & 38.15 & 36.07 & \nd{33.71} \\
  &  & \MambaSplit{} & 52.95 & 50.35 & 47.60 & 44.52 & 41.66 & 39.96 & \nd{38.22} & \nd{36.21} & 33.70 \\
  &  & \quad $+$ TTA & \nd{53.62} & \best{50.81} & \best{47.97} & \nd{44.84} & \nd{42.08} & \best{40.35} & \best{38.55} & \best{36.51} & \best{33.92} \\
\bottomrule
\end{tabular}
\end{adjustbox}
\end{table*}

\begin{table*}[p]\centering
\caption{\textbf{RI-PSNR at every mixing ratio: BioSR, HT-LIF24 and HT-T24.} As in \cref{tab:perratio-a}.}
\label{tab:perratio-b}
\footnotesize\setlength{\tabcolsep}{4.5pt}\renewcommand{\arraystretch}{0.92}
\begin{adjustbox}{max width=\textwidth}
\begin{tabular}{@{}c c l ccccccccc@{}}
\toprule
& & & \multicolumn{9}{c}{Channel-A weight $w$} \\
\cmidrule(l){4-12}
\textbf{Dataset} & \textbf{Ch.} & \textbf{Model} & 0.1 & 0.2 & 0.3 & 0.4 & 0.5 & 0.6 & 0.7 & 0.8 & 0.9 \\
\midrule
 \multirow{16}{*}{\rotatebox[origin=c]{90}{\textbf{BioSR}}} & \multirow{8}{*}{A} & U-Net & 22.02 & 26.80 & 30.22 & 32.39 & 33.98 & 34.97 & 35.51 & 35.89 & 36.71 \\
  &  & $\mu$Split & 23.23 & 26.90 & 30.21 & 32.37 & 33.72 & 34.65 & 35.34 & 36.04 & 37.72 \\
  &  & denoiSplit & 22.30 & 26.60 & 29.92 & 31.97 & 33.44 & 34.31 & 34.66 & 34.71 & 35.23 \\
  &  & MicroSplit & 23.54 & 27.34 & 30.68 & 32.82 & 34.42 & 35.77 & 36.86 & 37.78 & 39.24 \\
  &  & InDI & 23.26 & 27.50 & 30.43 & 32.13 & 33.06 & 33.49 & 33.69 & 34.11 & 35.24 \\
  &  & scSplit & 26.20 & 29.24 & 31.57 & 33.53 & 34.92 & 35.94 & 36.94 & 38.21 & 40.06 \\
  &  & \MambaSplit{} & \nd{28.46} & \nd{32.12} & \nd{34.51} & \nd{36.23} & \nd{37.80} & \nd{39.13} & \nd{40.43} & \nd{41.95} & \nd{44.12} \\
  &  & \quad $+$ TTA & \best{28.95} & \best{32.49} & \best{34.88} & \best{36.61} & \best{38.26} & \best{39.63} & \best{41.17} & \best{42.81} & \best{45.08} \\
\cmidrule(l){2-12}
  & \multirow{8}{*}{B} & U-Net & 39.51 & 38.06 & 37.41 & 35.96 & 33.90 & 31.22 & 28.12 & 24.81 & 21.65 \\
  &  & $\mu$Split & 41.69 & 38.75 & 37.55 & 35.82 & 33.47 & 30.79 & 27.87 & 24.66 & 21.59 \\
  &  & denoiSplit & 39.32 & 37.53 & 36.74 & 35.27 & 33.26 & 30.56 & 27.41 & 24.07 & 21.32 \\
  &  & MicroSplit & 40.48 & 38.61 & 37.74 & 36.26 & 34.29 & 32.03 & 29.38 & 26.17 & 22.34 \\
  &  & InDI & 37.91 & 37.50 & 36.89 & 35.81 & 34.06 & 31.58 & 28.68 & 25.60 & 22.43 \\
  &  & scSplit & 44.79 & 41.41 & 39.42 & 37.72 & 35.92 & 33.88 & 31.62 & 28.77 & 24.58 \\
  &  & \MambaSplit{} & \nd{46.88} & \nd{43.91} & \nd{41.59} & \nd{39.57} & \nd{37.76} & \nd{35.77} & \nd{33.57} & \nd{30.91} & \nd{26.84} \\
  &  & \quad $+$ TTA & \best{47.74} & \best{44.33} & \best{41.98} & \best{39.98} & \best{38.17} & \best{36.22} & \best{34.10} & \best{31.40} & \best{27.23} \\
\midrule
 \multirow{16}{*}{\rotatebox[origin=c]{90}{\textbf{HT-LIF24}}} & \multirow{8}{*}{A} & U-Net & 29.08 & 32.51 & 36.21 & 39.76 & 42.69 & 44.47 & 45.00 & 44.82 & 45.64 \\
  &  & $\mu$Split & 29.78 & 33.73 & 37.59 & 40.82 & \nd{43.22} & 44.60 & 45.20 & 45.25 & 45.66 \\
  &  & denoiSplit & 29.88 & 33.64 & 37.06 & 39.84 & 41.90 & 43.12 & 43.63 & 43.31 & 44.01 \\
  &  & MicroSplit & 30.01 & 33.93 & 37.65 & 40.65 & 42.80 & 44.20 & 44.66 & 44.63 & 44.28 \\
  &  & InDI & 32.36 & 36.54 & 39.59 & 41.05 & 41.12 & 40.43 & 39.60 & 38.88 & 38.92 \\
  &  & scSplit & 35.53 & 37.96 & 39.70 & 41.31 & 43.03 & 44.88 & 46.94 & 49.22 & \nd{52.37} \\
  &  & \MambaSplit{} & \nd{36.57} & \nd{38.60} & \nd{40.13} & \nd{41.54} & 43.15 & \nd{45.09} & \nd{47.23} & \nd{49.47} & 52.26 \\
  &  & \quad $+$ TTA & \best{36.88} & \best{38.87} & \best{40.45} & \best{41.92} & \best{43.60} & \best{45.58} & \best{47.82} & \best{50.15} & \best{53.16} \\
\cmidrule(l){2-12}
  & \multirow{8}{*}{B} & U-Net & 45.57 & 46.73 & 47.56 & 47.86 & 47.29 & 45.78 & 43.45 & 39.89 & 34.76 \\
  &  & $\mu$Split & 47.22 & 48.17 & 48.63 & 48.44 & 47.62 & 46.03 & 43.69 & 40.27 & 34.66 \\
  &  & denoiSplit & 45.93 & 45.96 & 45.66 & 45.11 & 44.35 & 43.14 & 41.48 & 38.88 & 34.06 \\
  &  & MicroSplit & 45.62 & 45.60 & 45.48 & 45.33 & 44.94 & 44.09 & 42.45 & 39.93 & 35.23 \\
  &  & InDI & 50.87 & 51.82 & 51.39 & 49.51 & 46.88 & 44.27 & 41.69 & 39.04 & 36.10 \\
  &  & scSplit & 57.06 & 54.03 & 51.43 & 49.42 & \nd{47.81} & \nd{46.31} & 44.60 & 42.37 & \nd{39.21} \\
  &  & \MambaSplit{} & \nd{58.39} & \nd{54.58} & \nd{51.81} & \nd{49.54} & 47.73 & 46.24 & \nd{44.69} & \nd{42.51} & 39.01 \\
  &  & \quad $+$ TTA & \best{59.15} & \best{55.03} & \best{52.21} & \best{49.96} & \best{48.17} & \best{46.68} & \best{45.16} & \best{42.96} & \best{39.39} \\
\midrule
 \multirow{16}{*}{\rotatebox[origin=c]{90}{\textbf{HT-T24}}} & \multirow{8}{*}{A} & U-Net & 32.63 & 34.83 & 37.21 & 39.40 & 41.06 & 41.62 & 40.96 & 39.76 & 38.63 \\
  &  & $\mu$Split & 32.16 & 34.33 & 36.67 & 38.96 & 40.78 & 41.44 & 40.83 & 39.68 & 38.62 \\
  &  & denoiSplit & 33.55 & 35.78 & 38.15 & 40.35 & 41.84 & 42.13 & 41.37 & 40.27 & 39.31 \\
  &  & MicroSplit & 32.62 & 34.86 & 37.12 & 39.27 & 40.86 & 41.41 & 40.95 & 39.98 & 39.10 \\
  &  & InDI & 32.13 & 33.74 & 35.83 & 38.37 & 41.01 & 43.11 & 43.63 & 42.56 & 41.09 \\
  &  & scSplit & \best{36.98} & \nd{39.25} & 40.88 & 41.98 & 42.92 & 44.09 & 45.81 & 48.25 & 50.86 \\
  &  & \MambaSplit{} & 36.51 & 39.24 & \nd{40.89} & \nd{42.02} & \nd{42.98} & \nd{44.89} & \nd{47.05} & \nd{49.77} & \nd{53.07} \\
  &  & \quad $+$ TTA & \nd{36.68} & \best{39.47} & \best{41.13} & \best{42.21} & \best{43.07} & \best{45.08} & \best{47.45} & \best{50.25} & \best{54.01} \\
\cmidrule(l){2-12}
  & \multirow{8}{*}{B} & U-Net & 35.57 & 36.01 & 36.64 & 36.38 & 34.67 & 31.85 & 28.59 & 25.30 & 22.25 \\
  &  & $\mu$Split & 35.38 & 35.39 & 35.61 & 35.16 & 33.65 & 31.24 & 28.37 & 25.33 & 22.30 \\
  &  & denoiSplit & 33.83 & 34.19 & 34.70 & 34.81 & 34.00 & 32.11 & 29.39 & 26.28 & 23.06 \\
  &  & MicroSplit & 33.84 & 33.76 & 33.98 & 33.90 & 32.98 & 31.19 & 28.74 & 25.82 & 22.85 \\
  &  & InDI & 32.48 & 32.76 & 32.89 & 32.88 & 32.30 & 31.45 & 29.81 & 27.24 & 24.19 \\
  &  & scSplit & 42.82 & 40.75 & 39.15 & 37.14 & 35.15 & 33.38 & 32.11 & 30.55 & 28.30 \\
  &  & \MambaSplit{} & \nd{45.72} & \nd{42.90} & \nd{40.52} & \nd{38.28} & \nd{35.98} & \nd{34.60} & \nd{33.27} & \nd{31.59} & \nd{28.84} \\
  &  & \quad $+$ TTA & \best{47.03} & \best{43.65} & \best{41.02} & \best{38.63} & \best{36.09} & \best{34.74} & \best{33.46} & \best{31.78} & \best{29.03} \\
\bottomrule
\end{tabular}
\end{adjustbox}
\end{table*}

\subsubsection*{S2.4\quad Weak and dominant components}\suplabel{sec:supp:qual}{S2.4}

\crefrange{fig:gallery-PaviaATN}{fig:gallery-HTT24} show one test frame per benchmark at both
$w=0.2$ and $w=0.8$, so each structure appears once as the weak and once as the dominant component.
\MambaSplit{}'s dominant-component crops are comparable to or better than those of the baselines, so its
weak-component gains are not obtained by sacrificing the dominant component.

\subsection*{S3\quad Severity adaptation and ablations}
\suplabel{sec:supp:severity}{S3}

\subsubsection*{S3.1\quad Ratio-estimation quality}
\suplabel{sec:supp:ratio}{S3.1}

Ratio estimation is accurate on four of the five datasets (\cref{fig:ratio-estimation-error}), with mean absolute errors between $0.017$ and $0.048$. BioSR is the exception. Replacing the predicted ratio $\hat{\alpha}$ with the true ratio at inference changes RI-PSNR by at most $0.25$\,dB on the other four datasets, but improves BioSR by $0.74$--$0.82$\,dB (\supref{sec:supp:conditioning}). Ratio-estimation error is therefore a minor limitation for most datasets, but has a larger effect on BioSR.

BioSR also shows that ratio MAE and reconstruction sensitivity need not
track each other. In a patch-size sweep at a fixed batch size of $5$ (not pixel-matched, unlike \cref{fig:patch-size}), as the training patch increases from $256$ to $512$
to $992$ pixels, per-tile ratio MAE changes from $0.063$ to $0.048$
to $0.051$, while the RI-PSNR gap between predicted- and true-ratio
inference decreases from $1.88$ to $0.79$ to $0.40$\,dB. The largest
crop thus reduces reconstruction sensitivity to ratio-estimation error
without further improving mean ratio accuracy.

Reconstruction quality in this sweep follows a different trend, peaking at $512$ pixels even when the true ratio is supplied. The $992$-pixel crop covers nearly the entire $1004\times1004$ BioSR frame and therefore provides very little crop diversity during training. Reduced training-view diversity is a plausible explanation for this non-monotonic reconstruction trend, although the present patch-size experiment changes context and crop diversity simultaneously and does not isolate their effects.

\figRatioEstimationError
\subsubsection*{S3.2\quad Frame pairing on BioSR}
\suplabel{sec:supp:pairing}{S3.2}

The released scSplit BioSR model uses index-matched training
frames, whereas \MambaSplit{} samples the two channels independently.
We retrain each method with the alternative sampling strategy,
keeping its remaining configuration unchanged (\cref{tab:pairing}).
Independent sampling improves RI-PSNR by $2.4$\,dB for scSplit
and $3.9$\,dB for \MambaSplit{} on the nine-ratio mean.
With index-matched frames, \MambaSplit{} trails scSplit by
$1.3$\,dB; with independent sampling, it leads by $0.2$\,dB.
A ratio-conditioned U-Net trained with the \MambaSplit{}
recipe achieves $37.0$\,dB, compared with $37.3$\,dB for
\MambaSplit{}.
Independent sampling is therefore an important part of the
training recipe and benefits both methods, although the size of
the gain differs substantially.

\begin{table}[!htb]
\centering
\small
\setlength{\tabcolsep}{6pt}
\caption{\textbf{Effect of training-frame sampling on BioSR.}
Single-pass RI-PSNR (dB) using the protocol of
\cref{tab:summed-severity-results} and final checkpoints.
Index-matched sampling uses the same frame index for both channels;
independent sampling draws their frames separately.
The U-Net uses the \MambaSplit{} training recipe
(\supref{sec:supp:unet}).
Best and second-best values are \best{bold} and
\nd{underlined}.}
\label{tab:pairing}

\begin{adjustbox}{max width=\columnwidth}
\begin{tabular}{@{}llrrrr@{}}
\toprule
\textbf{Model} & \textbf{Sampling}
& \textbf{Dominant} & \textbf{Balanced}
& \textbf{Weak} & \textbf{Mean} \\
\midrule
U-Net-38M
& Index-matched & 40.9 & 36.0 & 29.2 & 35.4 \\
& Independent   & \nd{42.4} & \nd{37.7} & 31.0 & 37.0 \\
\midrule
scSplit
& Index-matched & 40.1 & 35.3 & 28.7 & 34.7 \\
& Independent   & 42.1 & \best{37.9} & \best{31.3}
                & \nd{37.1} \\
\midrule
\MambaSplit{}
& Index-matched & 39.3 & 33.8 & 27.2 & 33.4 \\
& Independent   & \best{43.1} & \nd{37.7} & \nd{31.1}
                & \best{37.3} \\
\bottomrule
\end{tabular}
\end{adjustbox}
\end{table}
\subsubsection*{S3.3\quad What drives severity robustness}\suplabel{sec:supp:conditioning}{S3.3}

\begin{table*}[t]\centering\footnotesize\setlength{\tabcolsep}{2.6pt}
\caption{\textbf{Components of severity adaptation.} \MambaSplit{} gives absolute RI-PSNR (dB) in the
Dominant (D), Balanced (B) and Weak (W) regimes; the $\Delta$ rows give the change relative to it.
Fixed-ratio variants are trained only at $\alpha=0.5$; the true-$\alpha$ row uses the same
\MambaSplit{} checkpoint but supplies the true ratio at inference, and the TTA row averages its predictions over the eight
dihedral views (\supref{sec:supp:targets}). The TTA + true-$\alpha$ row combines both. The $\hat{\alpha}$-only row conditions FiLM on the predicted ratio itself instead of its four-term encoding (\supref{sec:supp:archcond}). The true-$\alpha$-only row feeds the true ratio to that same model at inference. The architecture-control row
replaces the shared decoder by one decoder per output channel.}
\label{tab:ratio-conditioning-ablation}
\begin{adjustbox}{max width=\textwidth}
\begin{tabular}{@{}l ccc ccc ccc ccc ccc@{}}
\toprule
& \multicolumn{3}{c}{\textbf{PaviaATN}} & \multicolumn{3}{c}{\textbf{Hagen}} & \multicolumn{3}{c}{\textbf{BioSR}}
& \multicolumn{3}{c}{\textbf{HT-LIF24}} & \multicolumn{3}{c}{\textbf{HT-T24}} \\
\cmidrule(lr){2-4}\cmidrule(lr){5-7}\cmidrule(lr){8-10}\cmidrule(lr){11-13}\cmidrule(l){14-16}
\textbf{Variant}
& D & B & W & D & B & W & D & B & W & D & B & W & D & B & W \\
\midrule
\MambaSplit{} & 40.13 & 30.22 & 28.09 & 47.20 & 40.52 & 36.27 & 43.15 & 37.71 & 31.07 & 52.29 & 45.55 & 40.25 & 46.50 & 39.79 & 35.06 \\
\midrule
$\Delta$ Fixed $\alpha=0.5$ + FiLM & $-$9.85 & $-$3.75 & $-$4.29 & $-$7.17 & $-$2.76 & $-$5.90 & $-$6.32 & $-$1.26 & $-$5.17 & $-$6.77 & $-$0.78 & $-$4.15 & $-$11.10 & $-$5.31 & $-$6.04 \\
$\Delta$ Fixed $\alpha=0.5$, no FiLM & $-$9.53 & $-$3.77 & $-$4.55 & $-$7.87 & $-$2.72 & $-$5.96 & $-$6.33 & $-$1.06 & $-$5.18 & $-$6.47 & $-$0.69 & $-$3.85 & $-$11.24 & $-$5.46 & $-$6.15 \\
$\Delta$ Uniform ratios, no FiLM & +0.05 & +0.05 & +0.03 & $-$0.87 & $-$0.42 & $-$0.26 & $-$0.38 & $-$0.30 & $-$0.14 & $-$0.45 & $-$0.33 & $-$0.25 & $-$0.14 & $-$0.18 & +0.03 \\
\midrule
$\Delta$ True $\alpha$ at inference & +0.20 & +0.12 & +0.04 & +0.08 & +0.01 & +0.04 & +0.81 & +0.82 & +0.74 & +0.10 & $-$0.01 & +0.04 & +0.25 & +0.01 & +0.08 \\
\midrule
$\Delta$ TTA at inference & +0.53 & +0.61 & +0.35 & +0.54 & +0.41 & +0.28 & +0.70 & +0.43 & +0.44 & +0.63 & +0.44 & +0.37 & +0.73 & +0.18 & +0.20 \\
$\Delta$ TTA + true $\alpha$ at inference & +0.63 & +0.70 & +0.40 & +0.64 & +0.41 & +0.32 & +1.58 & +1.39 & +1.15 & +0.73 & +0.42 & +0.40 & +0.90 & +0.17 & +0.30 \\
\midrule
$\Delta$ FiLM w/o $\varphi(\hat{\alpha})$ & $-$0.17 & $-$0.11 & $-$0.06 & $-$0.53 & $-$0.14 & $-$0.09 & $-$0.15 & $-$0.36 & $-$0.07 & $-$0.21 & $-$0.14 & $-$0.09 & +0.06 & $-$0.07 & +0.03 \\
$\Delta$ FiLM w/o $\varphi(\alpha)$ +  (True $\alpha$) & $-$0.02 & $-$0.07 & $-$0.05 & $-$0.45 & $-$0.13 & $-$0.05 & +0.80 & +0.75 & +0.66 & $-$0.11 & $-$0.15 & $-$0.07 & +0.28 & $-$0.08 & +0.08 \\
\midrule
$\Delta$ Separate decoders & +0.08 & $-$0.19 & $-$0.01 & +0.69 & +0.01 & +0.01 & $-$0.76 & $-$0.44 & $-$0.31 & $-$0.24 & $-$0.07 & +0.12 & $-$0.02 & $-$0.15 & +0.13 \\
\bottomrule
\end{tabular}
\end{adjustbox}
\end{table*}

\cref{tab:ratio-conditioning-ablation} expands the training-recipe ablation
(\cref{tab:recipe-ablation}) to all datasets. Restricting training to $\alpha=0.5$ causes losses
of up to approximately $11$\,dB away from the Balanced regime.
These ratios are out of distribution for fixed-$\alpha$ models, so
this ablation measures training-distribution coverage rather than
FiLM extrapolation. In contrast, removing FiLM from a model trained
across ratios changes RI-PSNR by less than $1$\,dB in every
dataset--regime pair.

Supplying the true ratio at inference gives small gains on four
datasets; BioSR is the exception, improving by $0.74$--$0.82$\,dB.
Broad exposure to mixing severities is therefore the main source
of robustness, with ratio conditioning providing a smaller
complementary benefit.

Separate decoders provide no consistent reconstruction gain while
increasing the parameter count from $36.41$\,M to $50.77$\,M and
model-forward time by $1.61\times$, so we retain the shared decoder.

\subsubsection*{S3.4\quad Sensitivity to the training-ratio distribution}\suplabel{sec:supp:dist}{S3.4}

We finally test whether severity-diverse training depends on the precise sampling distribution.
On HT-LIF24 and HT-T24, uniform, Gaussian and InDI-style sampling give similar performance,
whereas training only at $\alpha=0.5$ remains substantially worse
(\cref{tab:training-ratio-distribution}). All non-degenerate distributions cover the full severity range under
the monotonic $w(\alpha)$ mapping (\supref{sec:supp:axis}), but weight it differently:
Gaussian sampling emphasizes Balanced mixtures, while InDI-style
sampling adds extra mass at $\alpha=0.5$. Only fixed-$\alpha$
training is restricted to a single severity. Performance differs
by at most approximately $0.7$\,dB among the severity-diverse
distributions, indicating that broad coverage matters more than
its exact shape.

\begin{table}[!htbp]\centering\footnotesize\setlength{\tabcolsep}{2.6pt}
\caption{\textbf{Sensitivity to the training-ratio distribution.} RI-PSNR (dB) on HT-LIF24 and
HT-T24 in the Dominant (D), Balanced (B) and Weak (W) regimes; backbone, budget, seed and inference
are fixed, only the training-ratio distribution and FiLM change. Uniform samples $\alpha$ from $[0,1]$; $\mathcal{N}(\mu,\sigma)$ is a Gaussian. InDI-style sampling uses
$\alpha=0.5$ with probability $1/3$ and otherwise samples uniformly from $[0,1]$; Gaussian samples
are clipped to $[0,1]$. Best and second-best values are \best{bold} and \nd{underlined}.}
\label{tab:training-ratio-distribution}
\begin{tabular}{@{}l c ccc ccc@{}}
\toprule
& & \multicolumn{3}{c}{\textbf{HT-LIF24}} & \multicolumn{3}{c}{\textbf{HT-T24}} \\
\cmidrule(lr){3-5}\cmidrule(l){6-8}
\textbf{Training $\alpha$} & FiLM & D & B & W & D & B & W \\
\midrule
Fixed $0.5$ &  & 45.82 & 44.86 & 36.40 & 35.26 & 34.33 & 28.91 \\
Fixed $0.5$ & \checkmark & 45.52 & 44.77 & 36.10 & 35.40 & 34.48 & 29.01 \\
Uniform &  & 51.84 & 45.21 & 40.01 & 46.37 & 39.61 & \best{35.09} \\
Uniform (ours) & \checkmark & \best{52.29} & 45.55 & \best{40.25} & \nd{46.50} & 39.79 & 35.06 \\
$\mathcal{N}(0.5,0.15)$ &  & 51.63 & \nd{45.76} & 39.87 & 46.21 & \best{39.90} & 34.95 \\
$\mathcal{N}(0.5,0.15)$ & \checkmark & \nd{52.22} & \best{45.95} & 40.07 & 46.32 & \nd{39.87} & 34.90 \\
$\mathcal{N}(0.5,0.20)$ &  & 51.82 & 45.61 & 40.00 & 46.22 & 39.74 & 35.02 \\
InDI-style &  & 51.76 & 45.37 & 40.01 & 46.13 & 39.71 & 35.07 \\
InDI-style & \checkmark & 52.21 & 45.56 & \nd{40.20} & \best{46.62} & 39.76 & \nd{35.08} \\
\bottomrule
\end{tabular}
\end{table}

\subsection*{S4\quad Adaptation to experimentally acquired superpositions}
\suplabel{sec:supp:real}{S4}

\subsubsection*{S4.1\quad Transfer and adaptation}

Models trained on computationally summed inputs transfer imperfectly to experimentally acquired superpositions (\cref{tab:acquired-results}), so we adapt each method using the aligned acquired training pairs (\cref{tab:datasets}).

For \MambaSplit{}, training from scratch uses the unconditioned variant of \cref{sec:method:acquired}. Because acquired pairs provide no supervised mixing ratio, the LERP head, FiLM conditioning, and ratio-estimation loss are removed. The acquired input is normalized using training-set statistics, while the two component targets retain their channel-specific normalization.

Fine-tuning instead starts from the severity-aware summed-input model used in \cref{tab:summed-severity-results}. The pretrained LERP head is frozen and continues to predict a mixing descriptor from each acquired input, while the FiLM layers and reconstruction network remain trainable. Fine-tuning therefore preserves the representation learned from summed inputs while adapting the reconstruction network to the acquired domain.

\begin{table*}[!htbp]
\centering
\footnotesize
\setlength{\tabcolsep}{5pt}
\caption{\textbf{Per-channel results after training on acquired superpositions.}
RI-PSNR (dB) after training from scratch, fine-tuning for at most $20\%$ of the
training epochs, and fine-tuning to convergence
(\supref{sec:supp:finetune}). Best and second-best results within each
dataset and training stage are \best{bold} and \nd{underlined}.
$^{\ddagger}$Plateaued within the $20\%$ budget.
HT-T24 MicroSplit from-scratch results are from the published model.}
\label{tab:acquired-per-channel}

\begin{tabular}{@{}ll ccc ccc ccc@{}}
\toprule
& &
\multicolumn{3}{c}{\textbf{From scratch}} &
\multicolumn{3}{c}{\textbf{Fine-tuned, $\le$20\%}} &
\multicolumn{3}{c}{\textbf{Fine-tuned, converged}} \\
\cmidrule(lr){3-5}
\cmidrule(lr){6-8}
\cmidrule(l){9-11}
\textbf{Dataset} & \textbf{Model}
& A & B & Avg.
& A & B & Avg.
& A & B & Avg. \\
\midrule

HT-LIF24
& MicroSplit
& \best{42.07} & 45.38 & \nd{43.72}
& \nd{42.02} & 45.45 & 43.74
& 42.18 & 45.66 & 43.92 \\

& scSplit
& \nd{41.81} & \best{45.91} & \best{43.86}
& 42.01 & \nd{45.71} & \nd{43.86}
& \nd{42.23} & \best{46.31} & \nd{44.27} \\

& \MambaSplit{}
& 41.56 & \nd{45.59} & 43.57
& \best{42.34} & \best{46.22} & \best{44.28}
& \best{42.34} & \nd{46.22} & \best{44.28}$^{\ddagger}$ \\

\midrule

HT-T24
& MicroSplit
& 40.3 & 32.8 & 36.55
& 40.47 & 33.05 & 36.76
& 40.47 & 33.05 & 36.76$^{\ddagger}$ \\

& scSplit
& \best{40.89} & \best{33.13} & \best{37.01}
& \nd{40.94} & \nd{33.37} & \nd{37.16}
& \nd{41.05} & \nd{33.25} & \nd{37.15} \\

& \MambaSplit{}
& \nd{40.83} & \nd{33.10} & \nd{36.96}
& \best{41.07} & \best{33.61} & \best{37.34}
& \best{41.07} & \best{33.61} & \best{37.34}$^{\ddagger}$ \\

\bottomrule
\end{tabular}
\end{table*}

\subsubsection*{S4.2\quad Fine-tuning protocol and convergence}
\suplabel{sec:supp:finetune}{S4.2}

Each method is initialized from its own summed-input model and fine-tuned on the same acquired training split as its from-scratch run, using half the learning rate and at most $20\%$ of the original training epochs ($60$ of $299$ epochs for \MambaSplit{} and scSplit; $80$ of $400$ for MicroSplit). Checkpoints are selected by validation RI-PSNR and evaluated once on the test split.

\MambaSplit{} reaches its best validation checkpoint within this budget on both datasets, giving $44.28$\,dB on HT-LIF24 and $37.34$\,dB on HT-T24. Extending the baselines to convergence improves scSplit to $44.27$ and $37.15$\,dB and MicroSplit to $43.92$\,dB on HT-LIF24, while its HT-T24 run had already plateaued~(\cref{tab:acquired-per-channel}). Thus, longer fine-tuning does not change the ranking: \MambaSplit{} is effectively matched with scSplit on HT-LIF24 ($0.01$\,dB higher) and remains $0.19$\,dB higher on HT-T24.

\cref{tab:finetune-variants} compares alternative treatments of
the conditioning components. Freezing the ratio head, additionally
freezing FiLM, training all parameters, or removing both components
changes RI-PSNR by at most $0.05$\,dB on either dataset. The
fine-tuning result is therefore insensitive to these conditioning
choices on these datasets.

\begin{table*}[t]\centering\small\setlength{\tabcolsep}{4pt}
\caption{\textbf{Fine-tuning variants on acquired superpositions.} RI-PSNR (dB) of \MambaSplit{} for channels A and B and their average, single pass and with eight-view TTA (average only). All fine-tuning variants start from the same summed-input model, use half the learning rate and at most $20\%$ of the training epochs; they differ only in which conditioning parts are frozen or removed. The \emph{Ratio head frozen} row is the fine-tuned entry of \cref{tab:acquired-results}. Fine-tuned scSplit reaches $44.27$ (HT-LIF24) and $37.15$\,dB (HT-T24). Best and second-best values in each column are bold and underlined, respectively.}
\label{tab:finetune-variants}
\begin{tabular}{@{}l ccc c ccc c@{}}
\toprule
& \multicolumn{4}{c}{\textbf{HT-LIF24}} & \multicolumn{4}{c}{\textbf{HT-T24}} \\
\cmidrule(lr){2-5}\cmidrule(lr){6-9}
\textbf{Variant} & A & B & Avg & +TTA & A & B & Avg & +TTA \\
\midrule
Summed training, no fine-tuning
& 38.87 & 42.28 & 40.57 & 40.83
& 39.30 & 31.80 & 35.55 & 35.69 \\

Acquired training from scratch
& 41.56 & 45.59 & 43.57 & 44.06
& 40.83 & 33.10 & 36.96 & 37.16 \\
\midrule

Ratio head frozen (\cref{tab:acquired-results})
& \underline{42.34} & \underline{46.22} & \underline{44.28} & \underline{44.67}
& \textbf{41.07} & \underline{33.61} & \underline{37.34} & \textbf{37.48} \\

Ratio head and FiLM frozen
& 42.29 & 46.21 & 44.25 & 44.66
& \underline{41.06} & 33.58 & 37.32 & 37.38 \\

Nothing frozen
& 42.31 & 46.20 & 44.26 & 44.66
& 41.05 & 33.60 & 37.33 & 37.39 \\

FiLM and ratio head removed
& \textbf{42.35} & \textbf{46.25} & \textbf{44.30} & \textbf{44.68}
& \textbf{41.07} & \textbf{33.64} & \textbf{37.35} & \underline{37.47} \\
\bottomrule
\end{tabular}
\end{table*}

\cref{fig:acquired-qual-more} provides two additional acquired test examples per dataset after fine-tuning.

\figAcqQualMore
\subsection*{S5\quad Architecture and computational efficiency}

\subsubsection*{S5.1\quad U-Net comparison}
\suplabel{sec:supp:unet}{S5.1}

\begin{table*}[t]\centering\small\setlength{\tabcolsep}{4.5pt}
\caption{\textbf{U-Net comparison.}
RI-PSNR (dB) on HT-LIF24 and HT-T24 across Dominant (D),
Balanced (B), and Weak (W) regimes; Avg is the mean across
all six values.
All U-Nets are trained by us. ``Our recipe'' uses
uniformly sampled mixing ratios, $\alpha\sim\mathcal{U}[0,1]$,
and FiLM conditioning. The $-$FiLM and Vanilla rows report
RI-PSNR differences relative to the corresponding
``Our recipe'' model; Vanilla uses a fixed $\alpha=0.5$
without conditioning.
Parameters, runtime, and peak memory are reported as
absolute values. Inference is measured on a
$2720\times2720$ image using $1024\times1024$ tiles,
stride $256$, and batch size $4$ on the same $71$\,GB GPU
(\cref{tab:inference-cost}).
Bold indicates the best result among the two U-Nets
trained with our recipe and \MambaSplit{}.}
\label{tab:unet-conditioning}
\begin{tabular}{@{}l c ccc ccc c c c@{}}
\toprule
& & \multicolumn{3}{c}{\textbf{HT-LIF24}} & \multicolumn{3}{c}{\textbf{HT-T24}} & & & \\
\cmidrule(lr){3-5}\cmidrule(lr){6-8}
\textbf{Model} & \makecell{Par.\\(M)}
& D & B & W & D & B & W & Avg & \makecell{Runtime\\(s)} & \makecell{Mem.\\(GB)} \\
\midrule
U-Net-15M (Our recipe) & 14.79 & 51.96 & 45.76 & 40.22 & 45.86 & 39.15 & 34.49 & 42.91 & 3.86 & 5.00 \\
U-Net-15M (Our recipe): $-$FiLM & 14.67 & $-$0.26 & $-$0.43 & $-$0.32 & $-$0.53 & $-$0.10 & +0.12 & $-$0.25 & 3.79 & 5.00 \\
U-Net-15M (Vanilla) & 14.67 & $-$6.00 & $-$1.25 & $-$3.98 & $-$8.93 & $-$4.53 & $-$4.73 & $-$4.90 & 3.79 & 5.00 \\
\midrule
U-Net-38M (Our recipe) & 38.37 & \textbf{52.96} & \textbf{46.12} & \textbf{40.56} & 46.17 & 39.31 & 34.67 & \textbf{43.30} & 9.74 & 8.26 \\
U-Net-38M (Our recipe): $-$FiLM & 38.20 & $-$0.59 & $-$0.30 & $-$0.19 & $-$0.48 & $-$0.22 & +0.01 & $-$0.29 & 9.64 & 8.25 \\
U-Net-38M (Vanilla) & 38.20 & $-$7.83 & $-$1.15 & $-$4.60 & $-$9.22 & $-$4.27 & $-$4.65 & $-$5.29 & 9.64 & 8.25 \\
\midrule
\MambaSplit{} & 36.41 & 52.29 & 45.55 & 40.25 & \textbf{46.50} & \textbf{39.79} & \textbf{35.06} & 43.24 & \textbf{2.93} & \textbf{4.20} \\
\bottomrule
\end{tabular}
\end{table*}

To test whether the severity-training recipe depends on the Mamba
backbone, we train U-Nets of two capacities, with the same decoder design as MambaSplit, using the same summed-input construction and training-ratio distributions (\cref{tab:unet-conditioning}). Uniform ratio sampling provides
the dominant gain at both capacities, while FiLM gives a smaller
additional improvement. Severity robustness is therefore not
specific to Mamba.

Reconstruction accuracy is comparable: \MambaSplit{} is $0.52$\,dB
lower than U-Net-38M (Our recipe) on HT-LIF24 and $0.40$\,dB
higher on HT-T24. The clearer difference is computational: under
the same $1024\times1024$ tiling workload at batch size $4$, the U-Net requires
$3.32\times$ the model-forward time and about $2\times$ the
peak memory.
Since a U-Net trained with the same recipe and ratio conditioning reaches similar accuracy, most of the accuracy comes from the training recipe rather than from the backbone; \MambaSplit{}'s role is to make the large tiles of this recipe practical, as the fastest model in \cref{tab:unet-conditioning} at the same batch size and with faster frame processing at $1024$-pixel tiles than at smaller ones (\cref{tab:tile-speed}).

The U-Nets were trained with $16$-bit mixed precision and \MambaSplit{} with bfloat$16$, so sub-decibel reconstruction differences should be treated as qualitative.

\subsubsection*{S5.2\quad Inference accounting}
\suplabel{sec:supp:profiling}{S5.2}

Runtime values include model-forward computation only;
checkpoint loading, stitching, and metric evaluation are excluded.
All methods are timed in fp32 on the same
$71$\,GB GPU partition (one MIG slice of an NVIDIA H200).
Released baselines retain their native tile sizes,
posterior-sampling counts, reconstruction networks, and
test-time views.

For a PaviaATN frame, this corresponds to
100 tile evaluations for U-Net, 1,000 for denoiSplit,
352,800 for MicroSplit, and 2,100 for scSplit, comprising
2,000 reconstruction passes and 100 ratio-regressor passes.
\MambaSplit{} requires 64 tile evaluations.
These counts refer to tile evaluations, which may be
grouped into batched forward calls.
\Cref{tab:unet-conditioning} compares \MambaSplit{} with
U-Net under a matched $1024\times1024$ tiling workload
at a batch size of $4$.

The decoder-design control in
\cref{tab:ratio-conditioning-ablation} uses $36.41$\,M
parameters with the shared decoder and $50.77$\,M with
separate decoders; separate decoders also need $1.06\times$
the peak memory and $1.61\times$ the model-forward time
of the shared decoder for the same tiles, a ratio that does
not depend on the inference stride.

\cref{tab:batch-policy} reports time and peak memory at
batch size $1$ and at the largest batch that fits.
Batching reduces baseline inference time, most notably
for MicroSplit ($4.2\times$).
All $64$ \MambaSplit{} tiles fit into a single batched
forward call on the $71$\,GB GPU, but maximum batching
is slightly slower than processing tiles individually
($3.18$ versus $2.96$\,s) and uses substantially more peak
memory ($58.34$ versus $1.50$\,GB).

\begin{table}[!htb]\centering\small\setlength{\tabcolsep}{4pt}
\caption{\textbf{Inference time and memory under two batch policies.} One $2720\!\times\!2720$ PaviaATN frame on the $71$\,GB GPU of \cref{tab:inference-cost}, model-forward time only. Batch size $1$ for every method, and the largest batch that fits on the slice (used in \cref{tab:inference-cost}). Peak memory grows with the batch.}
\label{tab:batch-policy}
\begin{adjustbox}{max width=\columnwidth}
\begin{tabular}{@{}l r r r r r r@{}}
\toprule
& & \multicolumn{2}{c}{\textbf{One tile per pass}} & \multicolumn{3}{c}{\textbf{Largest batch}} \\
\cmidrule(lr){3-4}\cmidrule(l){5-7}
\textbf{Method} & \textbf{Passes} & \textbf{Time (s)} & \textbf{Mem. (GB)} & \textbf{Batch} & \textbf{Time (s)} & \textbf{Mem. (GB)} \\
\midrule
U-Net$^{\dagger}$ & 100 & 0.86 & 0.29 & 100 & 0.76 & 26.79 \\
denoiSplit & 1{,}000 & 28.48 & 0.75 & 64 & 24.85 & 47.36 \\
MicroSplit$^{\dagger}$ & 352{,}800 & 2196.14 & 0.06 & 1024 & 522.36 & 43.87 \\
scSplit$^{\dagger}$ & 2{,}100 & 24.85 & 0.30 & 128 & 18.69 & 30.41 \\
\midrule
\MambaSplit{} & 64 & 2.96 & 1.50 & 64 & 3.18 & 58.34 \\
\bottomrule
\end{tabular}
\end{adjustbox}
\end{table}

\cref{tab:tile-speed} compares tile sizes at the same relative
overlap, with the stride a quarter of the tile.
The default $1024$-pixel tile at stride $256$ processes
a frame $1.5\times$ faster than $512$-pixel tiles at stride
$128$ and $1.9\times$ faster than $256$-pixel tiles at
stride $64$, with the largest batch
($1.5\times$ and $2.3\times$ with one tile per pass).

\begin{table}[!htb]\centering\small\setlength{\tabcolsep}{4pt}
\caption{\textbf{Inference time and tile size.} Model-forward time for one $2720\!\times\!2720$ PaviaATN frame with one \MambaSplit{} model, trained on $1024$-pixel patches, tiled at $1024$, $512$ and $256$ pixels, with an inference stride of a quarter of the tile. Measured on the $71$\,GB GPU of \cref{tab:inference-cost}, one tile per pass and with the largest batch that fits (the two policies of \cref{tab:batch-policy}). Relative time uses the largest batch, with respect to the default $1024$-pixel tile.}
\label{tab:tile-speed}
\begin{adjustbox}{max width=\columnwidth}
\begin{tabular}{@{}r r r r r r@{}}
\toprule
\textbf{Tile} & \textbf{Stride} & \textbf{Tiles} & \makecell{\textbf{Time (s)}\\\textbf{one tile per pass}} & \makecell{\textbf{Time (s)}\\\textbf{largest batch}} & \textbf{Relative} \\
\midrule
1024 & 256 & 64 & 2.96 & 3.18 & $1.0\times$ \\
512 & 128 & 361 & 4.50 & 4.81 & $1.5\times$ \\
256 & 64 & 1600 & 6.86 & 6.17 & $1.9\times$ \\
\bottomrule
\end{tabular}
\end{adjustbox}
\end{table}
\subsection*{S6\quad Detection of tiling artifacts}
\suplabel{sec:supp:tiling}{S6}

\begin{table}[!t]\centering\footnotesize\setlength{\tabcolsep}{3.5pt}
\caption{\textbf{Tile-seam artifacts, Balanced regime.} SWITi permutation test; lower is better. FRT is the fraction of rejected equal-sample tests on each method's own seam grid; cFRT is the fraction of rejected tests on the common $256$-pixel seam grid (settings in \cref{tab:switi-protocol}). Best and second-best values, including TTA, are \best{bold} and \nd{underlined}. $^{\dagger}$Our unclipped retraining on PaviaATN and Hagen; other datasets use released models. On BioSR, \MambaSplit{} uses $512$-pixel tiles and a $128$-pixel grid.}
\label{tab:seam-balanced}
\begin{adjustbox}{max width=\columnwidth}
\begin{tabular}{@{}l c ccccc@{}}
\toprule
\textbf{Model} & Grid & PaviaATN & Hagen & BioSR & HT-LIF24 & HT-T24 \\
\midrule
\multicolumn{7}{@{}l}{\textit{FRT $\downarrow$}} \\
Ground truth & 32 & 0.054 & 0.047 & 0.037 & 0.046 & 0.049 \\
Ground truth & 128 & 0.053 & 0.043 & 0.040 & 0.037 & 0.049 \\
Ground truth & 256 & 0.050 & 0.039 & 0.044 & 0.032 & 0.047 \\
\cmidrule(l){2-7}
U-Net$^{\dagger}$ & 256 & \nd{0.100} & 0.217 & \best{0.136} & \best{0.056} & 0.165 \\
$\mu$Split & 32 & \best{0.086} & \best{0.072} & 0.443 & 0.206 & 0.315 \\
denoiSplit & 256 & 0.933 & 0.485 & \nd{0.158} & 0.500 & 0.659 \\
MicroSplit$^{\dagger}$ & 32 & 0.921 & 0.646 & 0.401 & 0.685 & 0.653 \\
scSplit$^{\dagger}$ & 256 & 0.270 & \nd{0.190} & 0.647 & 0.269 & 0.326 \\
\cmidrule(l){2-7}
\MambaSplit{} & 256 & 0.141 & 0.361 & 0.401 & 0.184 & \nd{0.164} \\
\quad $+$ TTA & 256 & 0.110 & 0.384 & 0.403 & \nd{0.140} & \best{0.141} \\
\midrule
\multicolumn{7}{@{}l}{\textit{cFRT $\downarrow$}} \\
Ground truth &  & 0.046 & 0.044 & 0.000 & 0.032 & 0.048 \\
\cmidrule(l){2-7}
U-Net$^{\dagger}$ &  & 0.291 & \nd{0.447} & \nd{0.342} & \best{0.083} & \best{0.380} \\
$\mu$Split &  & \best{0.174} & \best{0.125} & 0.608 & 0.334 & 0.745 \\
denoiSplit &  & 0.996 & 0.493 & \best{0.250} & 0.307 & 0.739 \\
MicroSplit$^{\dagger}$ &  & 0.996 & 0.579 & 0.425 & 0.395 & 0.710 \\
scSplit$^{\dagger}$ &  & 0.599 & 0.524 & 0.917 & 0.445 & 0.693 \\
\cmidrule(l){2-7}
\MambaSplit{} &  & 0.339 & 0.817 & 0.767 & 0.404 & 0.468 \\
\quad $+$ TTA &  & \nd{0.270} & 0.780 & 0.700 & \nd{0.304} & \nd{0.436} \\
\bottomrule
\end{tabular}
\end{adjustbox}
\end{table}

We use the \textit{Fraction of Rejected Tests} (FRT) introduced in~\cite{switi} to detect and quantify discontinuities at tiling seams in each method's stitched output.
FRT is derived from a per-tile, two-sample permutation test that compares one-step intensity gradients across a tile's seams with those in its immediate neighborhood.
The two samples are: $(i)$~the \emph{seam sample}, \ie, the gradient slice taken across each seam owned by the tile; $(ii)$~the \emph{control sample}, \ie, gradient strips drawn parallel to the seams, from both sides of each seam.
The null hypothesis is that the two samples come from the same distribution, \ie, that stitching introduces no excess discontinuity.
Since neighboring gradients are spatially correlated, the test permutes contiguous blocks of pixels along each slice between the two samples, rather than individual pixels.
The test uses the Jensen--Shannon divergence as the statistic $T$ and computes the empirical $p$-value as $p=(1+\#\{T_{\mathrm{null}}\ge T_{\mathrm{obs}}\})/(1+R)$ over $R$ permutations.
A test is \emph{rejected} if $p_{\mathrm{tile}}<0.05$, and FRT is the fraction of rejected tests in an image.
At the dataset level, we report FRT as the mean score over images.
In~\cref{tab:switi-protocol} we report the permutation-test settings used in our experiments.
\begin{table}[!htb]\centering\footnotesize\setlength{\tabcolsep}{4pt}\renewcommand{\arraystretch}{1.1}
\caption{\textbf{SWITi evaluation protocol.} Settings used for the tiling artifacts analysis in \cref{tab:seam-balanced} and \cref{fig:seam-vs-weight}.}
\label{tab:switi-protocol}
\begin{tabular}{@{}l p{0.70\columnwidth}@{}}
\toprule
\textbf{Setting} & \textbf{Value} \\
\midrule
Test statistic & Jensen--Shannon divergence on $32$-bin histograms \\
Block size & $3$ pixels \\
Control strip width & $2$ pixels \\
\# Permutations & $1{,}000$ \\
Significance & $p<0.05$ \\
\bottomrule
\end{tabular}
\end{table}

\begin{table}[!htbp]
\centering
\small
\setlength{\tabcolsep}{4pt}
\caption{\textbf{Effect of inference stride.}
\MambaSplit{} predictions are combined by uniform averaging
over overlapping tiles. FRT and cFRT are measured in the
Balanced regime (\cref{tab:seam-balanced}); ground-truth
scores are provided for reference. $\Delta$RI-PSNR denotes
the change in nine-ratio mean RI-PSNR (dB) relative to the
default stride of $256$ pixels ($128$ for BioSR). Rel.\ runtime is the runtime relative to the default stride (lower is faster), from the number of tile evaluations per frame.}
\label{tab:stride-blending}

\begin{adjustbox}{max width=\columnwidth}
\begin{tabular}{@{}c r cccc@{}}
\toprule
\textbf{Dataset} & \textbf{Stride} &
\textbf{FRT}\,$\downarrow$ &
\textbf{cFRT}\,$\downarrow$ &
$\Delta$\textbf{RI-PSNR} &
\makecell{\textbf{Rel.}\\\textbf{runtime}} \\
\midrule

\multirow{4}{*}{\makecell{PaviaATN\\(patch 1024)}}
& \textit{GT} & \textit{0.050} & \textit{0.046} & -- & -- \\
& 128  & 0.084 & 0.170 & $+0.04$ & $3.5\times$ \\
& 256  & 0.141 & 0.339 & ref. & $1\times$ \\
& 512  & 0.324 & 0.455 & $-0.27$ & $0.39\times$ \\
\midrule

\multirow{4}{*}{\makecell{Hagen\\(patch 1024)}}
& \textit{GT} & \textit{0.039} & \textit{0.044} & -- & -- \\
& 128  & 0.289 & 0.574 & $+0.01$ & $3.2\times$ \\
& 256  & 0.361 & 0.817 & ref. & $1\times$ \\
& 512  & 0.624 & 0.814 & $-0.10$ & $0.36\times$ \\
\midrule

\multirow{4}{*}{\makecell{HT-LIF24\\(patch 1024)}}
& \textit{GT} & \textit{0.032} & \textit{0.032} & -- & -- \\
& 128  & 0.110 & 0.214 & $+0.05$ & $2.3\times$ \\
& 256  & 0.184 & 0.404 & ref. & $1\times$ \\
& 512  & 0.433 & 0.486 & $-0.11$ & $0.56\times$ \\
\midrule

\multirow{4}{*}{\makecell{HT-T24\\(patch 1024)}}
& \textit{GT} & \textit{0.047} & \textit{0.048} & -- & -- \\
& 128  & 0.108 & 0.239 & $+0.23$ & $2.3\times$ \\
& 256  & 0.164 & 0.468 & ref. & $1\times$ \\
& 512  & 0.494 & 0.644 & $-0.66$ & $0.56\times$ \\
\midrule

\multirow{4}{*}{\makecell{BioSR\\(patch 512)}}
& \textit{GT} & \textit{0.040}
             & \textit{0.000} & -- & -- \\
& 64  & 0.298 & 0.317 & $+0.13$ & $3.2\times$ \\
& 128  & 0.401 & 0.767 & ref. & $1\times$ \\
& 256  & 0.679 & 0.992 & $-0.47$ & $0.36\times$ \\
\bottomrule
\end{tabular}
\end{adjustbox}

\end{table}
\figTilingArtefacts

Splitting methods use different tile sizes and overlaps at inference, so their tiling grids are not directly comparable.
We therefore report two FRT variants.
First, we evaluate FRT on \emph{each method's own tiling grid}, which reflects the seam structure that method actually leaves in the stitched output.
Second, we introduce \emph{common-grid FRT} (cFRT), which evaluates every method on a common lattice, defined as the coarsest grid shared across methods (here, a 256\,px stride), and probes whether stitching discontinuities persist at a scale shared by all methods.
\cref{fig:tiling-artefacts} (middle and bottom rows) provides an example of the grids used in the two FRT variants.
Ground-truth images are evaluated with the corresponding protocols and give rejection rates close to the nominal $0.05$ level.

The two FRT variants each require one adjustment to remain comparable across methods.
First, since test power scales with sample size, the per-method FRT assigns each test the same sample size across grids.
Each seam is split into contiguous 32\,px segments, and the segments at the same position along a tile's owned seams form one test.
Every test thus has the sample size of a 32\,px tile, and every seam pixel is used exactly once.
Second, the cFRT aligns the methods' tiling lattices to the common 256\,px grid by reflect-padding the first tile of each row and column in every stitched output during inference.

\figSeamQuality
FRT and cFRT only test whether tile seams stand out from their local neighborhood, not whether the predicted content is correct.
As a result, some baselines such as \UNet and \muSplit can obtain the lowest seam scores on several datasets despite considerably lower RI-PSNR, as noted in \cref{sec:exp:tiling}.
\cref{fig:seam-quality} shows an example on Hagen: \muSplit has considerably lower FRT and cFRT than \MambaSplit, yet its RI-PSNR is much lower (3.9\,dB lower on the frame and 5.5\,dB on the inset), showing that \MambaSplit better preserves the target structures.
FRT and cFRT should therefore be read together with RI-PSNR, not as standalone measures of reconstruction quality.
In the Balanced regime (\cref{tab:seam-balanced}), \MambaSplit{}'s rejection rates are moderate rather than the lowest: except for FRT on HT-T24, at least one baseline, most often \UNet or \muSplit, rejects fewer tests on every dataset and in both scores, and on Hagen \MambaSplit{} has the highest cFRT of all methods.

\cref{tab:stride-blending} varies the inference stride. Seam scores rise as the overlap between tiles shrinks, by $1.7$--$3\times$ from a quarter to half the tile size, and halving the overlap costs $0.1$--$0.7$\,dB RI-PSNR.

Finally, \cref{fig:seam-vs-weight} relates tiling artifacts to the difficulty of the splitting task by plotting FRT and cFRT against each output channel's mixing weight.
For most methods, rejection rates are highest in the Weak regime, where the channel contributes little to the mixture, and decrease as the mixing weight grows.
The decline is steepest for \MambaSplit and scSplit, which approach the chance level in the Dominant regime across several datasets.
Tiling artifacts thus grow with the difficulty of the task: when a structure is faint, independently processed tiles are more likely to disagree on how to assign its intensity, and these disagreements surface at the seams.
\figSeamVsWeight
\needspace{0.36\textheight}%
\begin{table}[H]\centering\small\setlength{\tabcolsep}{4pt}
\caption{\textbf{Tile seams vs.\ training patch size.} cFRT (Balanced regime, shared $256$-pixel grid; lower is better) and nine-ratio RI-PSNR (dB) of \MambaSplit{} models trained on $1024$-, $512$- or $256$-pixel patches with the same pixels per step (\cref{fig:patch-size}), each tiled at its own training patch and all inferred at stride $128$, so that all models of a row have the same seam lines; larger tiles therefore overlap more. Checkpoints as in \cref{fig:patch-size}. BioSR has no $1024$-pixel model.}
\label{tab:seam-patch}
\begin{tabular}{@{}l ccc ccc@{}}
\toprule
& \multicolumn{3}{c}{\textbf{cFRT}} & \multicolumn{3}{c}{\textbf{RI-PSNR}} \\
\cmidrule(lr){2-4}\cmidrule(l){5-7}
\textbf{Dataset} & $1024$ & $512$ & $256$ & $1024$ & $512$ & $256$ \\
\midrule
PaviaATN & \textbf{0.170} & 0.374 & 0.700 & \textbf{32.86} & 32.49 & 31.32 \\
Hagen    & \textbf{0.574} & 0.782 & 0.930 & \textbf{41.34} & 41.14 & 40.64 \\
HT-LIF24 & \textbf{0.214} & 0.384 & 0.665 & 46.08 & \textbf{46.15} & 45.63 \\
HT-T24   & \textbf{0.239} & 0.372 & 0.873 & \textbf{40.68} & 40.16 & 39.01 \\
BioSR    &  & \textbf{0.767} & 0.992 &  & \textbf{37.31} & 35.00 \\
\bottomrule
\end{tabular}
\end{table}

\cref{tab:seam-patch} varies the training patch instead. All its models are inferred at stride $128$, so each row compares the same seam lines. cFRT rises as the training patch shrinks on every dataset (Hagen $0.57$, $0.78$ and $0.93$ for $1024$-, $512$- and $256$-pixel patches; HT-LIF24 $0.21$, $0.38$ and $0.67$), and RI-PSNR falls on every dataset except HT-LIF24, where the $512$-pixel model is $0.07$\,dB above the $1024$-pixel model.

\subsection*{S7\quad Qualitative examples}
\suplabel{sec:supp:galleries}{S7}

Representative weak- and dominant-component reconstructions for all datasets are shown in \crefrange{fig:gallery-PaviaATN}{fig:gallery-HTT24}.

\figSynGalleries

\suppendpage

\end{document}